\documentstyle[aaspp4,epsf]{article}

\lefthead{Four-Year DMR Data}
\righthead{Bunn \& White}

\newcommand{\dH}{\delta_{\rm H}}
\newcommand{\hMpc}{\,h^{-1}{\rm Mpc}}
\newcommand{\npix}{{N_{\rm pix}}}
\newcommand{\rhat}{{\hat{\bf r}}}
\newcommand{\calm}{{\cal M}}
\newcommand{\half}{\hbox{$1\over 2$}}

\begin{document}

\title{The Four-Year COBE Normalization and Large-Scale Structure}
\author{Emory F. Bunn}
\affil{Department of Astronomy, University of California\\
Berkeley, CA 94720-7304}
\and
\author{Martin White}
\affil{Enrico Fermi Institute, University of Chicago\\
5640 S.~Ellis Ave, Chicago IL 60637}

\begin{abstract}
We present an analysis of the four-year data from the {\sl COBE} DMR
experiment.  We use a Karhunen-Lo\`eve expansion of the pixel data to
calculate the normalization and goodness-of-fit of a range of models of
structure formation.  This technique produces unbiased normalization
estimates and is capable of determining the normalization of any particular
model with a statistical uncertainty of 7\%.
We present a parameterization of the normalization and likelihood function
which summarizes our results for a wide range of models.  We use the
{\sl COBE} normalization to compute small-scale fluctuation amplitudes for
a variety of models and discuss the implications of these results for
theories of large-scale structure. 
\end{abstract}

\keywords{cosmology:theory --- cosmic microwave background --- large-scale
structure of universe}

\twocolumn

\section{Introduction} \label{sec:intro}

The {\sl COBE} DMR experiment has now been completed 
(\cite{Ben96,Goretal96,Hin96,Ban96}), 
and the four-year sky maps that were produced are the last word on
large-angle cosmic microwave background (CMB) anisotropies that we are
likely to have for some time.  The impact of these data on models of
structure formation has been immense.  By providing a theoretically clean
measure of the mass fluctuations in the linear regime, the {\sl COBE} data
has allowed for the first time a $\la10\%$ determination of the amplitude
of the power spectrum of cosmological models.  This normalization supersedes
the previous best method of normalization based on the abundance of clusters;
alternatively, the two normalizations together constrain one extra combination
of cosmological parameters.
The combination of cluster abundances and the {\sl COBE} normalization on
large-scales has definitively ruled out the ``standard'' Cold Dark Matter
(CDM) model with adiabatic, scale-invariant initial fluctuations.
For more discussion on the impact of the {\sl COBE} data on large-scale
structure theories, see (\cite{WhiSco}).

The {\sl COBE} data determine the amplitude of the fluctuation spectrum at
large scales with a statistical error of $7\%$.
In order to fully exploit this high-precision determination one must go
beyond the simple Sachs-Wolfe (\cite{SacWol}) approximation to the
large-angle anisotropy spectrum and a simple summary of the {\sl COBE}
data such as the rms fluctuation or the correlation function.
In this paper we present normalizations of several popular CDM based models
of structure formation by performing a maximum-likelihood fit of numerically
computed anisotropy spectra directly to the {\sl COBE} pixel data.  Tests
using Monte-Carlo data sets indicate that our maximum-likelihood estimates
are unbiased and that the maximum-likelihood points provide ``good'' fits to
the data.
We quote our results for both the radiation and matter power spectra at large
scales, which in any {\it particular\/} model have a definite relationship.
We also discuss how one computes the normalization on smaller scales from our
results.

The outline of the paper is as follows.  In \S\ref{sec:likelihood} we discuss
the method used to perform the maximum-likelihood fits to the {\sl COBE} data,
which is based on the Karhunen-Lo\`eve (KL) transform.
In this section we also present results of our Monte-Carlo simulations to test
for bias in the fitting method.
In \S\ref{sec:freq} we present two different frequentist methods for using the
{\sl COBE} data to constrain models, and we contrast these methods with the
Bayesian analysis used in the rest of the paper. 
The normalization of the radiation power spectra for models in the CDM family
is considered in \S\ref{sec:radiation}, where we also provide fitting functions
for a class of models with approximately quadratic anisotropy spectra.  For
open (OCDM) and cosmological constant ($\Lambda$CDM) cold dark matter models
we give likelihood functions vs.~$\Omega_0$ from the {\sl COBE} data.
In \S\ref{sec:matter} we use the results of \S\ref{sec:radiation} to find the
amplitude of the matter power spectrum for a wide range of CDM models.
After describing some implications of the {\sl COBE} normalization for models
of large-scale structure in \S\ref{sec:lss}, we present our conclusions in
\S\ref{sec:conclusions}.

\section{Likelihood Analysis} \label{sec:likelihood}

\subsection{Notation}

The normalizations and likelihoods presented in this paper are based
on an analysis of the four-year {\sl COBE} DMR sky maps (\cite{Ben96}).
As discussed in (\cite{BunScoWhi,Ban94,WhiBun}) there is much more
information in the {\sl COBE} DMR data than simply the rms fluctuation,
so detailed fitting of a theory to the pixel data is necessary to obtain all
of the information available from this data set.
We analyze the data by performing a Karhunen-Lo\`eve (KL) transform,
{\it i.e.}, expanding the data in a set of modes that optimally retain signal
and throw away noise.  This method is also known as the signal-to-noise
eigenmode analysis, or optimal subspace filtering.
For further information on this method, see
(\cite{Bond,BunSug,BunScoWhi,WhiBun,Bunn,VogSza}).

Let us start by describing the likelihood analysis of the {\sl COBE} DMR
pixel data.
As usual, we expand the CMB temperature anisotropy $\Delta T/T$
in spherical harmonics:
\begin{equation}
{\Delta T\over T}(\hat{\bf r})=\sum_{\ell,m}a_{\ell m}Y_{\ell m}(\hat{\bf r}).
\label{eqn:ylmexpansion}
\end{equation}
The {\sl COBE} data have been shown to be consistent with Gaussian
fluctuations (\cite{Kog96}) such as are predicted by the simplest
inflationary theories and by most defect theories (on large scales).
We will only consider Gaussian theories in this paper, hence each
$a_{\ell m}$ will be an independent Gaussian random variable of zero mean.
Thus all of the cosmological information in the CMB is contained in the
variances 
\begin{equation}
C_\ell\equiv \langle|a_{\ell m}|^2\rangle.
\end{equation}
The quantities $\{C_\ell\}$ are collectively referred to as the angular
power spectrum.  
Our goal in analyzing the DMR data is to test hypotheses about the angular
power spectrum.

We work with a single DMR sky map consisting of a weighted average of
the two 53 GHz maps and the two 90 GHz maps, in the
ecliptic pixelization.  We weight the maps
according to the inverse of the noise variance in each pixel, in order
to have minimal noise in the resulting map.  Under the
assumption of Gaussian statistics, this is mathematically
equivalent to performing a multi-map likelihood analysis.
(We do not use the 31 GHz maps for fear of Galactic contamination; in any
case, the noise levels in these maps are high enough that little information
is lost by not using them.)
We remove all pixels that lie within the ``custom cut'' described in
(\cite{Ben96}), leaving $\npix=3890$ pixels.  In addition,
we remove a best-fit monopole and dipole.  This last step is formally
unnecessary, since our likelihood analysis is explicitly forced to
be insensitive to the monopole and dipole.

Let $d_i$ be the temperature measurement corresponding to the $i$th
pixel, and let $\vec d$ be the $\npix$-dimensional vector
$(d_1,\ldots,d_\npix)$.
(Throughout this paper, we will denote vectors in real three-dimensional space
by boldface type, and vectors in other spaces such as ``pixel space'' by
arrows.)
The datum $d_i$ contains contributions from both signal and noise:
\begin{equation}
d_i=({\Delta T\over T}\star W)(\rhat_i)+n_i.
\end{equation}
Here $W(\rhat,\rhat')$ represents the window function of the experiment
(\cite{Wri94}), the star denotes convolution, $\rhat_i$ is the location
on the sky of the $i$th pixel and $n_i$ is a random variable representing
the noise.
The DMR noise is approximately Gaussian, and noise correlations from pixel
to pixel are weak (\cite{Line}).  The noise covariance matrix 
\begin{equation}
{\bf N}\equiv \langle \vec n\vec n^T\rangle
\end{equation}
is therefore approximately diagonal.

Since the underlying, theoretically uncorrelated variables are specified
in $\ell$ space we wish to write $\vec{d}$ in terms of the $a_{\ell m}$.
For notational convenience we will denote a pair of indices $(\ell m)$ by
a single Greek index such as $\mu$.  The correspondence between the two is
$\mu=\ell(\ell+1)+m+1$ so that $\mu$ ranges from 1 to $\infty$ as $\ell$
and $m$ vary over their entire allowed ranges.
We introduce an $\npix\times\infty$ matrix $\bf Y$ whose elements are
$Y_{i\mu}=Y_\mu(\rhat_i)$.
The most natural way to represent the window function is by a diagonal
matrix $W_{\mu\nu}=W_\ell\delta_{\mu\nu}$, where $\ell$ is the index
corresponding to $\mu$ and $W_\ell$ is the Legendre polynomial expansion
of the window function.  Then we can write
\begin{equation}
\vec d={\bf Y}{\bf W}\vec a+\vec n.
\end{equation}
Here $\vec a$ and $\vec n$ are the vectors whose components are $a_\mu$
and $n_i$ respectively.

The covariance matrix, $C_{\mu\nu}$, of $\vec{a}$ is diagonal, and its nonzero
elements are those of the angular power spectrum $C_\ell$:
\begin{equation}
C_{\mu\nu}\equiv\left\langle a_\mu a_\nu\right\rangle
  = C_\ell\ \delta_{\mu\nu},
\end{equation}
where $\ell$ is the index corresponding to $\mu$ as before.  We know the
covariance matrices of $\vec a$ and $\vec n$, and we assume that the signal
and noise are uncorrelated: $\left\langle a_\mu n_i\right\rangle=0$.
We can therefore write down the covariance matrix of the data vector $\vec d$:
\begin{equation}
\label{eqn:dcovar1}
{\bf M}\equiv\langle\vec d \vec d^T\rangle=
{\bf Y} \left( {\bf W}{\bf C}{\bf W} \right) {\bf Y}^T+{\bf N}.
\end{equation}
The matrix ${\bf W}{\bf C} {\bf W}$ is the beam-smoothed angular power
spectrum, and recall ${\bf W}^{T}={\bf W}$.

Since we are assuming that both the signal and noise are Gaussian, we know
the entire probability distribution function of the data vector $\vec d$:
\begin{equation}
\label{eqn:likely1}
f(\vec d \,|\, {\bf C})={1\over (2\pi)^{\npix/2}\det^{1/2} {\bf M}}
\exp\left(-\half \vec d^T{\bf M}^{-1}\vec d\right),
\end{equation}

We will regard $f$ as a function of ${\bf C}$ rather than as a function of the 
data $\vec d$: after all, $\vec d$ is fixed and ${\bf C}$ is unknown.
In this context, $f$ is called the {\it likelihood\/} and is denoted
$L({\bf C})$.  The maximum-likelihood estimate of some set of parameters
$\vec q$ (which may of course be a set containing only one element), is that
for which $L({\bf C}(\vec q))=L(\vec{q})$ is maximized.
The Bayesian credible region, assuming a uniform prior for the parameters,
is a volume $V$ bounded by a surface of constant $L$ such that
\begin{equation}
\label{eqn:credible}
\int_V L(\vec q)\,d\vec q=c\int L(\vec q)\,d\vec q,
\end{equation}
where $c$ is the confidence level associated with the region, and the
integral on the right-hand side extends over all of $\vec q$ space.
Note that this procedure is completely general, and in no way depends
upon the assumption of Gaussian statistics.  Of course, if we do not
make the assumption of Gaussianity, then Eq.~(\ref{eqn:likely1}) is not
valid, and we need to replace it with something else.

These prescriptions tell us how to perform our parameter estimates.
Unfortunately, we have no guarantee that these estimates will be unbiased.
In the case of Gaussian statistics, parameter estimates are asymptotically
unbiased; however, since we deal with finite sample sizes, this guarantee
does not apply.
If we are concerned about bias, we therefore have no recourse but to perform
Monte Carlo simulations to test for it, as we discuss in \S\ref{sec:tests}.

\subsection{The Karhunen-Lo\`eve Transform}

The full likelihood function is awkward to compute, since it involves
inverting an $\npix\times\npix$ matrix.
This has been done for a small family of models using the two-year DMR
data (\cite{TegBun}), but if we want to look at a larger family of models,
we need to find a way to ``compress'' the data before computing the
likelihood.
This is the purpose of the KL transform.
The idea is to (linearly) project the data vector onto a subspace of smaller
dimension, and compute likelihoods based on the projected data vector.
We attempt to choose the subspace in such a way that the likelihood function
based on the new ``compressed'' data vector contains almost all of the useful
information in the original likelihood function.

To be specific, suppose that our
goal is to estimate the normalization of the power spectrum.  Then
the optimal $D$-dimensional subspace to choose turns out to be that spanned
by the $D$ solutions $\vec\alpha_a$ with the largest eigenvalues
of the following eigenvalue equation:
\begin{equation}
\label{eqn:eigen}
{\bf M}_{\rm sig}\vec\alpha_a=\lambda_a{\bf M}\vec\alpha_a,
\end{equation}
where ${\bf M}_{\rm sig}\equiv {\bf YWCWY}^T$ is the signal part of
the covariance matrix
(\cite{Bond,BunSug,BunScoWhi,WhiBun,Bunn,VogSza}).

We should warn the reader that there are two definitions of the eigenvalues
in popular use.  For one choice the eigenvalues are in units of
signal-to-noise and range from 0 to infinity.  The other, used in this
paper, has the eigenvalues in the range 0 to 1.  The two are related by the
mapping $\lambda_a'=\lambda_a/(\lambda_a-1)$ with $\lambda_a$ as above.

Of course, we don't know $\bf M$ {\it a priori\/}, since we don't know the
power spectrum.
(If we did, there would be no need to do a likelihood analysis!).
We must therefore choose a ``fiducial'' power spectrum $C_\ell$ in order to
compute the eigenmodes.  We use a flat power spectrum
$\ell(\ell+1)C_\ell=$constant
as our fiducial power spectrum for the entire likelihood analysis, although
as we will describe below we have performed tests to show that our results
do not depend significantly on this choice.

\begin{figure}[t]
\centerline{\epsfxsize=3in\epsfbox{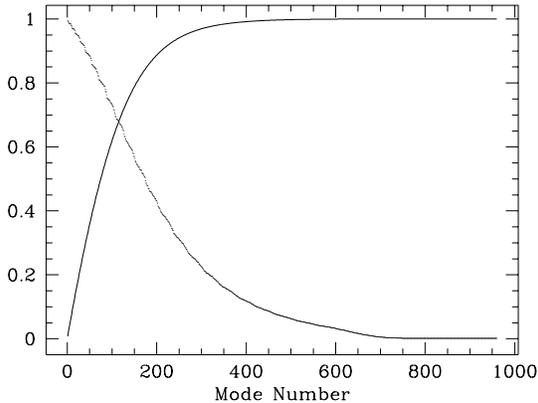}}
\caption{The points show the eigenvalues associated with the KL expansion of
the four-year {\sl COBE} data, sorted in decreasing order.  The solid curve
is a running sum of the squares of these eigenvalues, normalized
to a final value of 1.}
\label{fig:eigvals}
\end{figure}

The signal-to-noise eigenvalues are shown in Fig.~\ref{fig:eigvals}, where
it can be seen that the first $\sim400$ contain significant signal.  
The accuracy with which we can measure power spectrum normalizations
is determined by the sharpness of the peak of the likelihood function,
that is, by the quantity $-\langle L''(q)\rangle$, where $q$ represents
the normalization and the derivatives are evaluated at the peak of the
likelihood function.  This quantity is proportional to $\sum_a\lambda_a^2$,
which is therefore
the 
relevant figure of merit to describe our ability to reject incorrect models.
This quantity is also plotted in Fig.~\ref{fig:eigvals}.

\begin{figure*}[t]
\centerline{\epsfxsize=6in\epsfbox{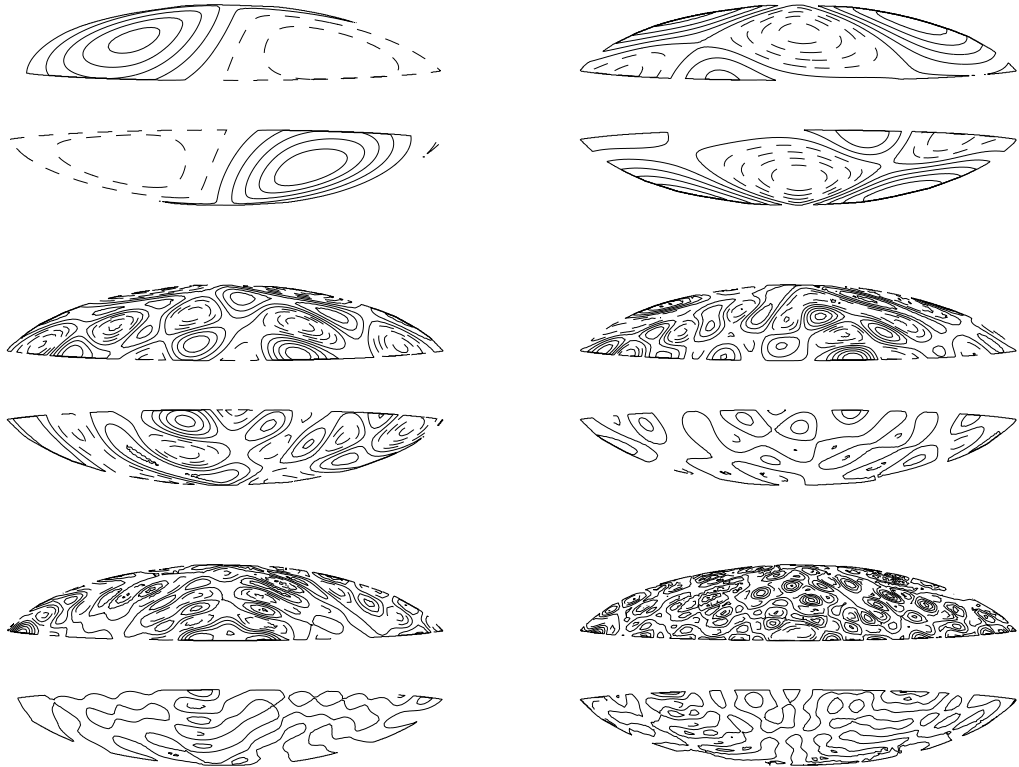}}
\caption{Aitoff-projected contour plots are shown of eigenmodes
1, 10, 50, 100, 200, running from left-to-right, top-to-bottom.
}
\label{fig:eigvecs}
\end{figure*}

\begin{figure}[t]
\centerline{\epsfxsize=3in\epsfbox{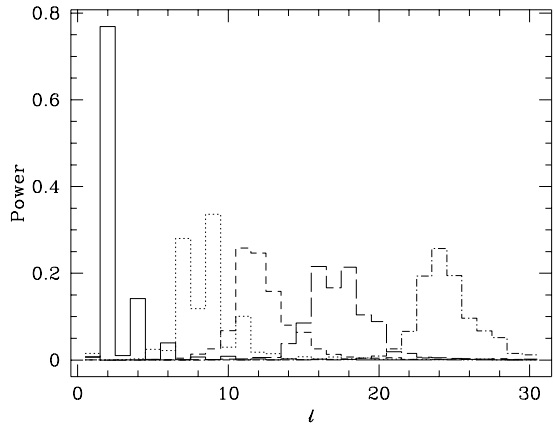}}
\caption{Histograms of the mean-square power as a function of $\ell$
are shown for various eigenmodes.
The modes plotted are numbers $a=1$, 50, 100, 200, and 400, from left
to right.}
\label{fig:eigells}
\end{figure}

Contour plots of a sample of eigenvectors in Aitoff projection are shown in 
Fig.~\ref{fig:eigvecs}.  The multipoles probed by these eigenvectors are
shown in Fig.~\ref{fig:eigells}.  Many of the eigenmodes show a surprising
lack of symmetry between the upper and lower hemispheres; mode 400,
for example, shows much more structure in the upper hemisphere than in
the lower.  In every such case, there is a nearly degenerate eigenmode
that is a approximately a reflection of the given mode.  For example,
mode 399 looks very much like mode 400, but with most of its structure
concentrated in the lower hemisphere.

Once we have chosen our subspace, we can project the data vector down onto
this subspace and compute likelihoods in terms of the projected vector.
However, before we can use these likelihoods we need to take proper account
of the monopole and dipole.  When we remove a best-fit monopole and 
dipole from the data, we are inadvertently removing some contribution from 
the other
modes, since incomplete sky coverage breaks the orthogonality of the spherical
harmonics.  From a Bayesian point of view, the natural way to correct for this
is to marginalize over the monopole and dipole.  That is, we should replace
$L$ by $\int L\ da_{00}da_{1-1}da_{10}da_{11}$, 
where $a_{00}$ is the monopole and $a_{1m}$ are the three dipole components.
All four of these quantities are unknown, so we marginalize over them
by letting them range over all possible values and integrating.  
Fortunately, the variation of $L$ with these four quantities has a simple
Gaussian form, and so the integral can be done analytically.
This procedure turns out to be mathematically equivalent to
forcing the eigenmodes to be orthogonal to the monopole and dipole,
{\it e.g.}, by Gram-Schmidt orthogonalization, as is done by
G\'orski et al.~(1996a).

In previous analyses of the DMR data, the quadrupole has frequently been
excluded along with the monopole and dipole.  Although we will give a few
quadrupole-excluded results below, most of our results will include the
information contained in the quadrupole.  Ever since
the initial detection of anisotropy in the one-year COBE maps, it has
been known that the COBE quadrupole moment is lower than one would
expect from ``preferred'' cosmological models ({\it i.e.}, those with
roughly scale-invariant power spectra) normalized to the modes with
$\ell\ge 3$.  Since we know {\it a priori\/} that the COBE quadrupole
is anomalous (compared to our theoretical expectations), we should
be quite hesitant to throw it away: in throwing away data that is known
{\it a priori\/} to be discordant, we run the risk of biased data editing.
In the absence of compelling evidence of quadrupolar contamination, we
therefore regard it as more prudent to retain the quadrupole.

There are many ways to estimate the quadrupole from the DMR data,
and in general they are not equivalent.  (In the absence of complete
sky coverage, there is no way to estimate a particular multipole
without contamination from other multipoles.)  If we estimate
the five components of the quadrupole by simply integrating
over the observed part of the sky,
\begin{equation}
b_{2m}={4\pi\over N_{\rm pix}}\sum_{i=1}^{N_{\rm pix}}
d_iY_{2m}(\hat r_i),
\end{equation}
and estimate the quadrupole by computing $\hat Q^2=(1/4\pi)\sum_m|\,b_{2m}|^2$
and subtracting off noise bias, we find that $\hat Q=5.9\,\mu\rm K$ for the
four-year DMR data.  Using the same estimator, we find quadrupoles of
$8.9\,\mu\rm K$ and $6.6\,\mu\rm K$ for the one-year and two-year data sets.
To decide whether this quantity is inconsistent with a particular model,
we need to compare it with simulations.  We find that $\hat Q$ is low enough
to be inconsistent with a flat Harrison-Zel'dovich power spectrum at about the
$98\%$ confidence level.  One should be reluctant to rule out any models on
this basis, however: we knew {\it a priori\/} that the quadrupole was low, and
confidence levels based on a subset of data that is known {\it a priori\/} to
be discordant can be misleading.

\subsection{Tests of our Method} \label{sec:tests}

To define our subspace we truncate the eigenmode expansion at $500$ vectors,
as we find that increasing the number of modes makes little difference to the
results.
For example, the estimate of the quadrupole normalization 
$Q\equiv\sqrt{5C_2/4\pi}$ for a flat
Sachs-Wolfe $n=1$ spectrum is $Q=18.73\pm 1.25\,\mu\rm K$ using 500
modes\footnote{With $500$ vectors the typical time to evaluate the likelihood
function at one point is $40$ seconds on a Sparc-10.}.
If we increase the number of modes to 700, we find $Q=18.67\pm
1.26\,\mu\rm K$.  Increasing the number of modes also fails to increase
significantly our sensitivity to the shape of the power spectrum.  For
example, using 500 modes our determination of the spectral index
$n$ (for a pure Sachs-Wolfe spectrum) is $n=1.184\pm 0.282$; using
700 modes this result becomes $n=1.177\pm 0.279$. 

Removing the quadrupole information from the fit (by marginalizing
over the quadrupole as well as the monopole and dipole) increased
the normalization to $Q=19.6$ for $n=1$, and changes the best-fit
value of $n$ from 1.2 to 1.0.

In order to assess the sensitivity of our results to the choice of
fiducial power spectrum, we computed some likelihoods using an
$n=1.5$ Sachs-Wolfe power spectrum (as opposed to the flat $n=1$ spectrum)
as our fiducial power spectrum.  The results are virtually identical.
For example, the maximum-likelihood estimate of $Q$ for a flat $n=1$
power spectrum, computed with an $n=1.5$ fiducial power spectrum,
is $18.74\,\mu$K.
Furthermore, the maximum-likelihood point in the $(n,Q)$ plane is
$(1.2,16.22\,\mu\rm K)$ for either a flat or an $n=1.5$ fiducial
power spectrum.

As mentioned before, we have no {\it a priori\/} guarantee that our
maximum-likelihood parameter estimates will be unbiased.
We have therefore performed Monte Carlo simulations to test for bias.
We generated 564 simulated sky maps with a pure $n=1$ Sachs-Wolfe power
spectrum and a normalization of $Q=19\,\rm \mu K$, and we determined the
maximum-likelihood normalization for each.
The average estimated normalization was $Q=18.98\,\mu\rm K$, with a
standard deviation of $\sigma=1.22\,\mu\rm K$.  Note that this standard
deviation is a frequentist estimate of the error in our estimate
of the normalization, which can be compared with the Bayesian error
estimate of $1.25\,\mu\rm K$ given above.

In the above test, the input power spectrum was the same as the fiducial power
spectrum.  Since this might not be a fair test, we also performed simulations
with an $n=1.5$ Sachs-Wolfe power spectrum and a scale-invariant
``standard CDM'' spectrum.  
For the $n=1.5$ spectrum, the input normalization was
$Q=13\,\mu$K, and the average estimated normalization was $12.99\,\mu$K with
a standard deviation of $0.83\,\mu$K.  
In the simulations of a standard CDM spectrum, we used an input
normalization of $Q=18\,\mu$K; in this case, the mean estimated
normalization was $18.03\,\mu$K with a standard deviation of $1.16\,\mu$K.
Based on these results, we are confident that there is no significant bias
in our normalization estimates.

We also tested for bias in our estimate of the slope $n$ of the power
spectrum.  We made simulated maps with Sachs-Wolfe input spectra
with parameters $(n,Q)=(1,19\,\mu\rm K)$ and $(n,Q)=(1.5,13\,\mu\rm K)$
as before, and found the maximum-likelihood point in the $(n,Q)$ plane
for each.  We found that the average estimate of $n$ in the two cases
was $1.02$ and $1.49$, with standard deviations of $0.26$ and $0.27$.

\begin{figure}[t]
\centerline{\epsfxsize=3.5in\epsfbox{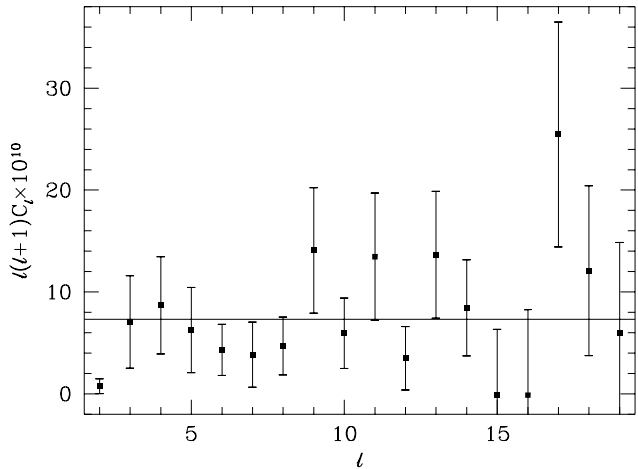}}
\caption{The maximum-likelihood power spectrum obtained by letting all
$C_\ell$'s with $2\le\ell\le 19$ vary freely is shown, together with a
$Q=19\,\mu\rm K$ flat Sachs-Wolfe power spectrum.  The error bars are
the standard errors determined by approximating the likelihood function
as a Gaussian near the peak; there are significant correlations between
the errors.}
\label{fig:bestfit}
\end{figure}

%Compare to COBE group.
Our results for pure Sachs-Wolfe spectra agree well with those of
(\cite{Goretal96}), though an exact comparison is not possible since we
use a combination of the maps which excludes the 31GHz map and weights
the 53GHz and 90GHz maps differently.  Furthermore, the KL
transform and the orthogonalized spherical harmonic
technique of G\'orski et al.~give
give slightly different weights to the various spherical
harmonic modes.  We therefore would not expect perfect quantitative
agreement between our results even if we used identical maps.
We would of course expect any discrepancies to be well within
the errors, though, and in fact we find that this is the case.

G\'orski et al.~(1996) quote a maximum likelihood point of
$(n,Q)=(1.22,15.9\mu$K) and a best fitting scale-invariant normalization
of $Q=18.7\mu$K if they include the quadrupole and use the ``coadded''
ecliptic maps with the custom cut (the closest to our procedure).
Our maximum likelihood point is $(n,Q)=(1.2,16.2\mu$K) and our best fitting
scale-invariant normalization $Q=18.7\mu$K.  The difference in likelihood
between $(1.2,16.2\mu$K) and $(1.22,15.9\mu$K) in our calculation is only
3\%, so this agreement is very good.
For the quadrupole {\it excluded\/} results, (\cite{Goretal96}) find the
most likely point is $(n,Q)=(1.00,19.1\mu$K) while we find
$(n,Q)=(1.00,19.6\mu$K).  Again the best fitting point of (\cite{Goretal96})
is only slightly less likely than our maximum likelihood point and
{\it well\/} within our $1\sigma$ error ellipse.

The best-fitting $C_\ell$ spectrum is shown in Fig.~\ref{fig:bestfit}.
To determine this spectrum, we let all $C_\ell$'s for $2\le \ell \le 19$
vary independently until the likelihood was maximized.  A flat power
spectrum with $Q=19\,\rm\mu K$ is shown for comparison.
The error bars shown in this figure are standard errors determined by
approximating the likelihood near the peak as a Gaussian.
The standard errors are then the square roots of the diagonal elements
of the covariance matrix of this Gaussian.  Error bars determined in this
way should be viewed with extreme caution.  First, the likelihood is not
very well approximated by a Gaussian: on the contrary, it is strongly
skew-positive at low $\ell$.  Second, these standard errors contain no
information about correlations between the errors.
These correlations are largest for pairs of modes whose $\ell$-values
differ by 2.  
Coupling  between modes with $\Delta\ell=1$ are weak because the data
has approximate reflection symmetry.
The normalized correlation coefficient
${\rm Cov}(C_\ell,C_{\ell'})/\sqrt{{\rm Var}(C_\ell)
{\rm Var}(C_{\ell'})}$ is typically $\sim 15\%$ at low $\ell$
for $\Delta\ell=2$ and decreases to $\sim 5\%$ for larger $\ell$.
The correlations are typically $\sim 5\%$ for $\Delta\ell=4$
over the whole range in $\ell$.
The deceptively small error bar on the estimate of $C_2$
is largely due to the failure of the Gaussian approximation
for the likelihood, although the 15\% anticorrelation
between $C_2$ and $C_4$ also plays a role.

\subsection{Filtering the Data}

\begin{figure*}[t]
\centerline{\epsfxsize=3in\epsfbox{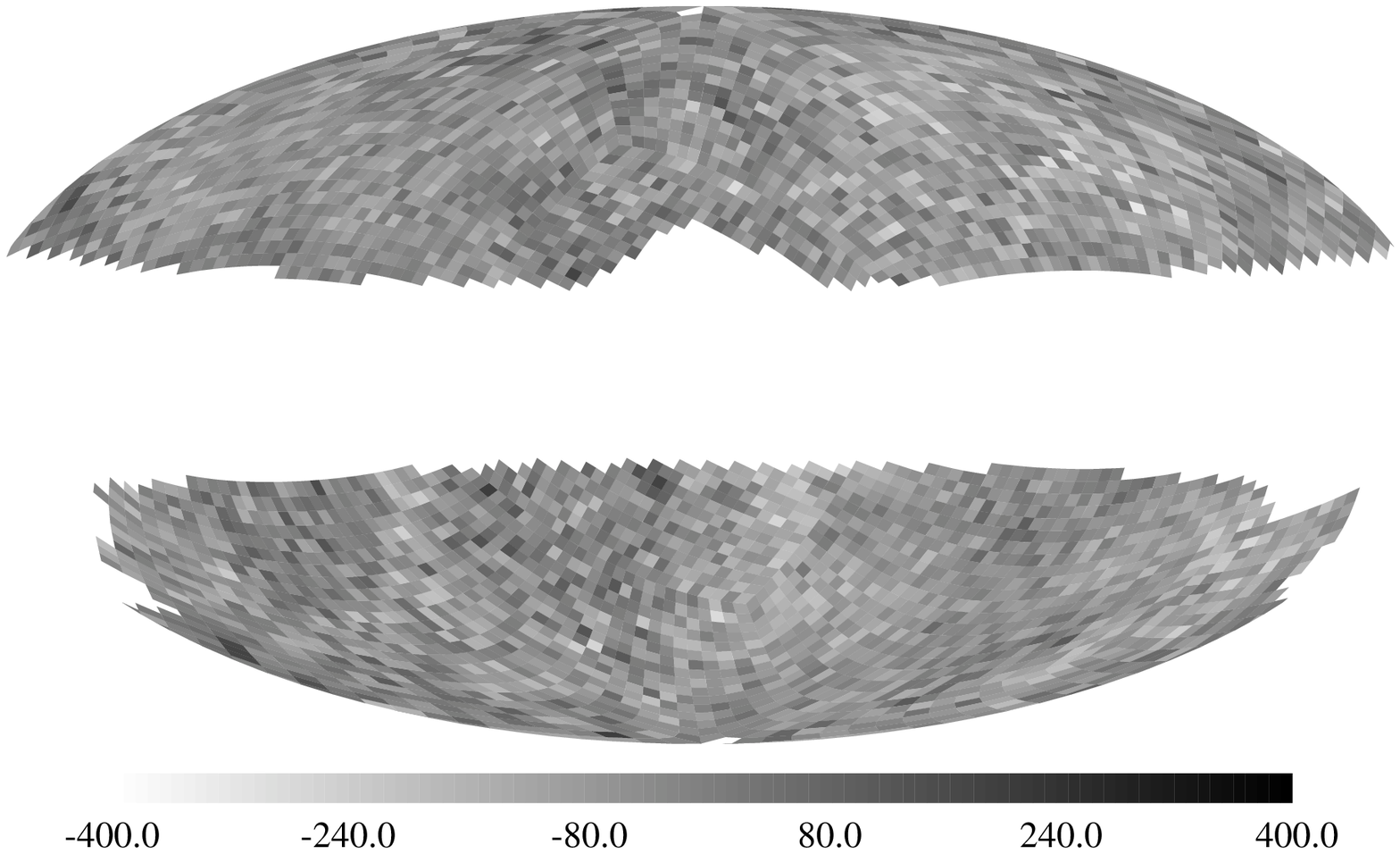}}
\centerline{\epsfxsize=3in\epsfbox{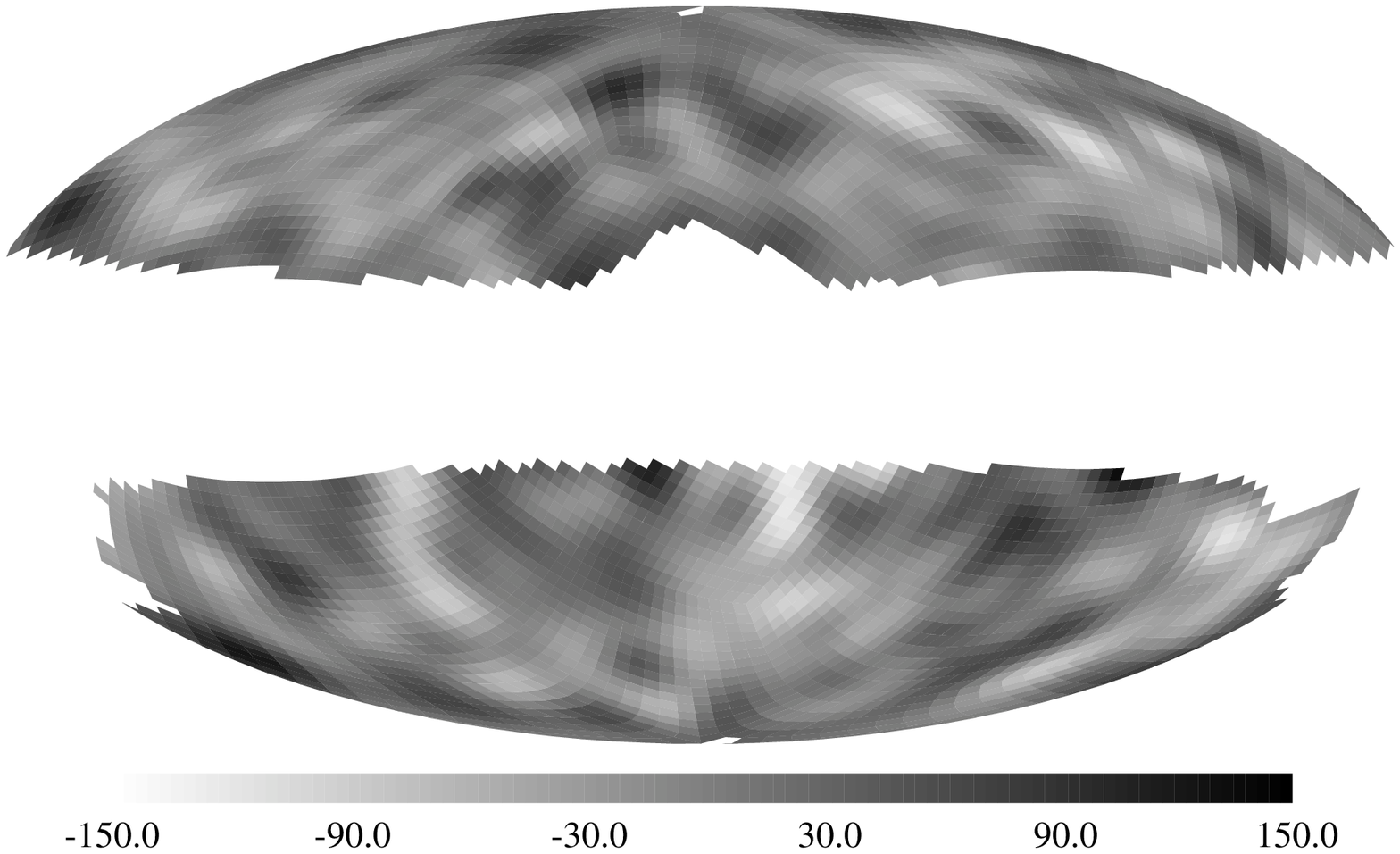}}
\centerline{\epsfxsize=3in\epsfbox{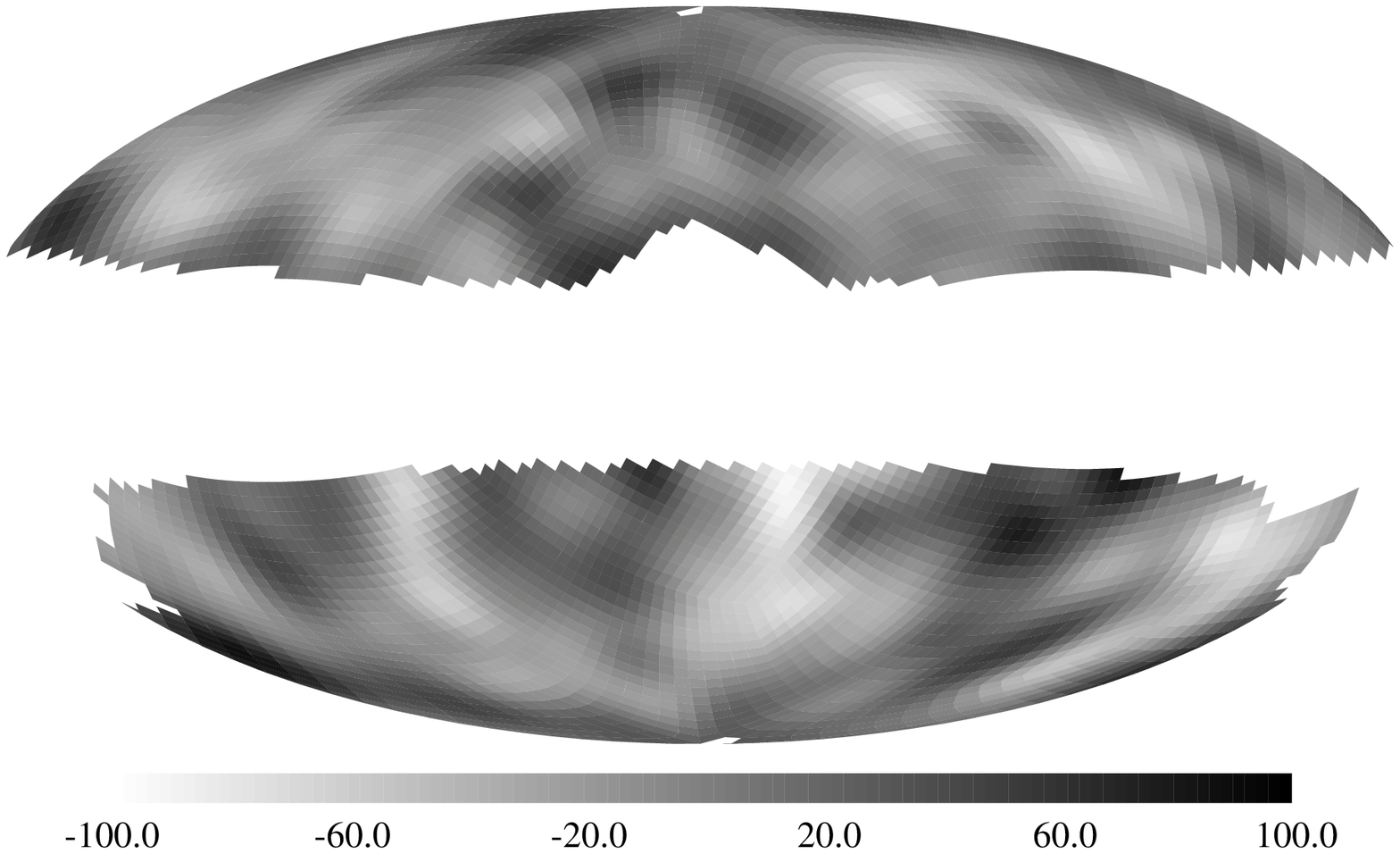}}
\caption{
The four-year {\sl COBE} DMR data are plotted in Aitoff projection.
The top panel shows the raw pixel data.  The middle panel is the
result of filtering the data by removing all but the first 400
signal-to-noise eigenmodes, and the lower panel shows the result
of applying a Wiener filter.}
\label{fig:klmaps}
\end{figure*}

The KL expansion is essentially a technique for optimally retaining signal
and throwing away noise in a data set.  As one might expect, it is closely
related to similar methods, such as the Wiener filter (\cite{BunHofSilk}).
Assuming that the statistical properties of the signal and noise are known,
applying the Wiener filter to a map preferentially removes noise and leaves
signal.  It is the optimal linear filter which can be constructed for this
purpose.

%Since the Wiener filter is designed for optimal reconstruction of the sky map
%(including phase information), whereas the KL transform is designed for
%efficient estimation of the power spectrum, the two are not identical.
Wiener filtering is particularly easy to perform
in the KL basis.
Specifically, the amplitude of the $a$th eigenmode of the Wiener-filtered sky
map is obtained simply by multiplying the amplitude of the same mode of the
raw map by the eigenvalue $\lambda_a$.  In Fig.~\ref{fig:klmaps} we show the
result of Wiener-filtering the DMR data in this way, as well as a map obtained
by simply truncating the KL expansion after 400 modes.

\section{Frequentist limits} \label{sec:freq}

The limits and error bars we have quoted so far in this paper have been based
upon Bayesian rather than frequentist techniques.
This has become standard practice in CMB data analysis, largely because
Bayesian results are often easier to compute than frequentist ones.
For a Bayesian, the likelihood function is all that is needed, whereas to find
the boundaries of a frequentist confidence interval one also needs to know the
theoretical probability distribution of some goodness-of-fit statistic.
These probability distributions can often only be computed by time-consuming
Monte Carlo simulations.

Although there is nothing intrinsically wrong with Bayesian methods, there are
certain circumstances in which a frequentist approach is to be preferred.
First, Bayesian estimates may sometimes depend too strongly on the assumed
prior, making interpretation of the results controversial.
As a rule, this is a serious problem only when the parameter being estimated
is weakly constrained by the data (e.g.~\cite{BunWhiSreSco}).
Second, although Bayesian techniques supply relative probabilities of different
models, they are in general unable to determine the intrinsic goodness of fit
of a single model.
As we saw above, we are able to find the most probable values of parameters
like the normalization $Q$ and the spectral index $n$; however, Bayesian
methods are unable to tell us whether or not the {\it best\/} fit is actually
a {\it good\/} fit.

It is therefore of interest to consider ways of applying frequentist techniques
to the {\sl COBE} data.  (We have in fact mentioned some frequentist error
estimates above, when we discussed the Monte Carlo simulations we performed to
test for bias in our maximum-likelihood parameter estimates.)
We begin by reminding the reader of the general approach a frequentist takes
to hypothesis testing.

Frequentist statistical analyses are always based on some
{\it goodness-of fit statistic\/} $\eta$.
The statistic should be chosen in such a way that it takes on low values when
the data fit the model well and large values when the fit is poor.  For a
particular hypothesis, we compute the probability distribution of $\eta$, and
we say that the hypothesis is ruled out with significance $s$ if
$P(\eta>\hat\eta)<s$, where $\hat\eta$ is the value found in the actual data.
For such an analysis to have high power, it is obviously essential to choose
$\eta$ wisely.

One common choice of goodness-of-fit statistic is the
{\it likelihood ratio}.
To illustrate this choice, let us consider a family of spatially flat, CDM
models with a cosmological constant ($\Lambda$CDM) and $n=1$. 
These models are characterized on {\sl COBE} scales by only two parameters,
the normalization $Q$ and the density $\Omega_0=1-\Omega_\Lambda$.
If we wish to test the hypothesis that $\Omega_0$ takes on some specific
value $\Omega_0^*$, we choose as a goodness-of-fit statistic
\begin{equation}
\eta = {\max_{Q,\Omega_0}L(Q,\Omega_0)\over
\max_QL(Q,\Omega_0^*)}.
\label{eq:likerat}
\end{equation}
If $\Omega_0$ is far from $\Omega_0^*$, then we expect this ratio to be large.
In order to decide whether a particular value of $\Omega_0^*$ is allowed by
the data, we compute the probability distribution of $\eta$ by making many
simulated {\sl COBE} sky maps with a power spectrum corresponding to
$\Omega_0=\Omega_0^*$ and computing the likelihood ratio (\ref{eq:likerat})
for each one.  This time-consuming process must be repeated for each value of
$\Omega_0^*$ we are interested in.

Upon performing this procedure for $\Omega_0^*=0.1$, we find that the
likelihood ratio is 3.16.  Only 10\% of simulated maps give a larger value,
and we therefore conclude that this particular model is ruled out with a
significance of 10\% (or at 90\% confidence).  By way of comparison, we shall
see below that a Bayesian analysis of the same family of models says that a
95\% credible lower limit on $\Omega_0$ is 0.13.  

Note that this family of models is an example of a situation where the
Bayesian limit will have significant prior dependence, since as
Fig.~\ref{fig:like_om} shows, the likelihood is large over a significant
fraction of the allowed parameter space.
A frequentist limit is therefore quite useful.  Unfortunately, frequentist
limits based on likelihood ratios are extremely expensive to compute, since
the probability distribution of $\eta$ must be recomputed by Monte Carlo
simulation for each point in parameter space.

The use of likelihood ratios to set frequentist limits alleviates only one of
the two disadvantages of the Bayesian approach: we have ``removed'' the prior
dependence, but the likelihood ratio still gives no information about intrinsic
goodness of fit: even if the data are an intrinsically poor fit to the whole
family of models, there will still be a value of $\Omega_0^*$ for which the
likelihood will peak and $\eta$ will take on its lowest possible value.
In (\cite{WhiBun}) we proposed a different goodness-of-fit statistic to resolve
this problem, and we will now describe a slight variant of this statistic.

Let $\vec x$ be the vector of amplitudes in our eigenmode expansion.  For a
particular theoretical model, we can compute the theoretical covariance matrix
$\calm\equiv\langle\vec x\vec x^T\rangle$, and then
transform to a new basis:
\begin{equation}
\vec y=\calm^{-1/2}\vec x.
\end{equation}
Any convenient square root of the covariance matrix may be chosen; we
recommend Cholesky decomposition.  The advantage of changing basis in this
way is that the predicted covariance matrix of $\vec y$ is the identity
matrix.  That is, each $y_i$ is predicted to be an independent unit Gaussian
random variable.  We shall see that this can save us a great deal of effort.

Next, we sort the $\{y_i\}$ from largest to smallest angular scale probed
(by computing the mean value of $\ell$ in a spherical harmonic expansion)
and sum $y_i^2$ over bins of some size $K$:
\begin{equation}
z_a=\sum_{i=(a-1)K+1}^{aK}y_i^2,
\end{equation}
where $1\le a\le N/K$ and $N$ is the total number of modes.
If our theoretical model is correct, then each $z_a$ should be
$\chi_K^2$-distributed and should therefore have expectation value $K$.
If our model is incorrect, then the dispersion of $z_a$ about this expectation
value should be larger.  We therefore propose as a goodness-of-fit statistic
the variance,
\begin{equation}
\eta={1\over 2N}\sum_{a=1}^{N/K} (z_a-K)^2.
\label{eq:eta}
\end{equation}
The prefactor $1/2N$ is arbitrary; we chose it so that the expectation value
of $\eta$ is one when the model is correct.

One great advantage of this statistic is that its probability distribution
is model-independent: $\eta$ is simply the average of $N/K$ independent
$\chi_K^2$-distributed random variables.  We need to compute this probability
distribution only once.  Furthermore, unlike the likelihood ratio, this
statistic is in principle capable of testing the possibility that
the whole family of models under consideration are poor fits to the data.

The statistic (\ref{eq:eta}) has two parameters: the number of modes
$N$ and the bin size $K$.  We experimented extensively with simulated
data sets to find the values of those statistics that maximized the
power to reject incorrect models.\footnote{ It is of course crucial to
perform such experiments on simulated data sets before looking at the
real data; choosing a statistic based on its behavior when applied to
the real data is a statistical {\it faux pas}.}  We made
simulated sky maps with a flat $\Omega_0=1$ power spectrum and tried
to find values of the parameters that maximized our ability to reject
an $\Omega_0=0.1$ model.  We found that the greatest power was
achieved when the number of bins was fairly small; we chose $N=200$
and $K=40$, so that the number of bins is only five.  Unfortunately,
the power of this statistic proves to be quite low: we are typically
able to reject the low-$\Omega_0$ model with significances of only
$\sim 20\%$.  As we have seen, the likelihood ratio allows
significantly greater power, rejecting the same model at 
a significance of $10\%$.

It is in general difficult to tell {\it a priori\/} whether a particular
goodness-of-fit statistic will be a powerful discriminator among
a given class of models; the statistic $\eta$ does not perform
as well in this regard as one might have hoped.
Although this statistic has low power, it is perhaps nonetheless
worth quoting some results based on it, since it is capable of
supplying intrinsic goodness-of-fit values for individual models
(a task for which the likelihood-ratio is ill-suited).
We find that, for the Harrison-Zel'dovich $\Lambda$CDM models
we have been considering in this section, the significance
ranges from 29\% when $\Omega_0=1$ to 17\% when $\Omega_0=0.1$,
with higher significances denoting better fits.  The fact that
all of these significance levels are acceptable is somewhat
reassuring: in principle we could have discovered that the whole
class of models fit the data at an unacceptably poor
level, in which case estimating parameters by means of likelihood
ratios or Bayesian methods would be a suspect procedure.

\section{Normalization of the Anisotropy Spectrum} \label{sec:radiation}

We now turn to the problem of quoting the best-fitting amplitude for various
models.  We break this into two parts: determining the best-fitting amplitude
of the {\it radiation\/} power spectrum and the best-fitting amplitude of the
{\it matter\/} power spectrum.
We discuss the first problem in this section.
Since the ratio of the radiation to the matter power spectrum is an output of
the model under consideration, the second follows from the first and will
be considered in \S\ref{sec:matter}.

The inadequacy of using only the RMS fluctuation in the map to determine the
amplitude of a given model from has been discussed by
(\cite{BunScoWhi,Ban94,WhiBun}).
As an example, normalizing a scale-invariant Sachs-Wolfe spectrum
(Eq.~\ref{eqn:swcl}) to the RMS anisotropy smoothed to $10^{\circ}$,
$\sigma_{10}=29\pm1\mu$K, one would obtain $10^{11}C_{10}=0.43\pm0.03$.
This is 30\% ($\sim2\sigma$) lower than obtained from a fit to the full data
set.  However, while the data clearly cannot be summarized by one number,
keeping the information from each and every pixel is redundant for the
theories we are considering.
Since fitting theories directly to the data is quite time consuming, in what
follows we will quote results which allow a simple but accurate method for
normalizing a given radiation spectrum to the full {\sl COBE} data.

\begin{figure}[t]
\centerline{\epsfxsize=3in\epsfbox{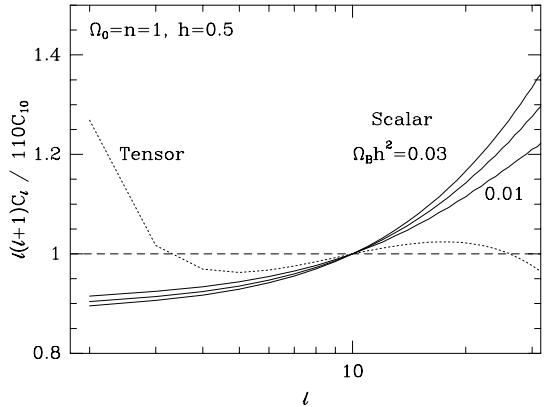}}
\caption{The angular power spectra $\ell(\ell+1)C_\ell$ for CDM models with
$\Omega_{\rm B}h^2=0.01$, 0.02, 0.03.  The tensor power spectrum is also shown.
All models are arbitrarily normalized at $\ell=10$.
Note that the curves differ significantly, even at low $\ell$, from the flat
dashed curve, which represents a pure Sachs-Wolfe spectrum.}
\label{fig:SW}
\end{figure}

\begin{figure}[t]
\centerline{\epsfxsize=3in\epsfbox{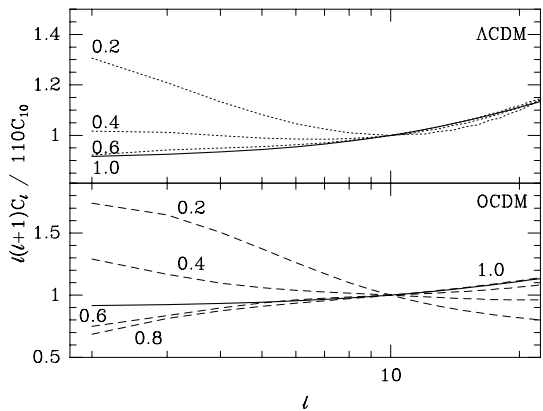}}
\caption{The shapes of the angular power spectra are plotted
for various CDM models.  The upper panel shows spatially flat
models with $\Omega_0+\Omega_\Lambda=1$, and the lower panel shows
models with $\Omega_\Lambda=0$.  All models have $n=1$ and are labeled
with the appropriate value of $\Omega_0$.}
\label{fig:cl_omega}
\end{figure}

For most models of structure formation the angular power spectrum at low $\ell$
contains little structure.  If we could simply parameterize the shapes of the
spectra of interest, we could pre-compute and tabulate the normalizations from
the {\sl COBE} data.  Which types of spectra should we investigate?
As can be seen in Figs.~\ref{fig:SW} and \ref{fig:cl_omega}, the radiation
power spectrum for a cold dark matter model is not that of a pure Sachs-Wolfe
spectrum of gravitational potential perturbations
(\cite{AbbWis,BonEfs,WSS,Pee93}):
\begin{equation}
C_\ell \propto
  {\Gamma(3-n)\Gamma\left(\ell+{n-1\over 2}\right)\over
  \Gamma^2\left({4-n\over 2}\right)\Gamma\left(\ell+{5-n\over 2}\right)} .
\label{eqn:swcl}
\end{equation}
For standard CDM the effect of including the full $C_\ell$ spectrum (that
is adding the integrated Sachs-Wolfe and velocity contributions) is to
reduce the best-fitting normalization by $\sim10\%$ at $C_2$, which gives
an indication of the curvature of the spectrum.
Including a full numerical calculation of the radiation spectrum also
introduces dependences on the cosmological parameters.
All except for $\Omega_0$ and $n$ are small, as discussed in
\S\ref{sec:matter}.

We follow (\cite{WhiBun}) in parameterizing the radiation power spectrum in
terms of quadratics.  Specifically, if we write
\begin{equation}
  D(x) = \ell(\ell+1)C_\ell \qquad {\rm with}\ x=\log_{10}\ell\quad ,
\end{equation}
then we provide the normalization for quadratic $D(x)$ using the methods
outlined in \S\ref{sec:likelihood}.

To normalize a theory then, one computes the radiation power spectrum for that
theory, finds the closest quadratic approximation to it over the range
relevant to {\sl COBE} (roughly $\ell=2$ to $30$) and reads off the
normalization which we provide for that quadratic.  This works quite well for
a large range of theories.
For example, for open and $\Lambda$CDM models with $0.2\le\Omega_0\le1$ and
$0.8\le n\le 1.1$ we find at most a 1\% difference between normalizing to a
quadratic approximation\footnote{Specifically we use the
best-fit quadratic to the $C_\ell$ between $\ell=3$ and $20$ with points
weighted equally in $\log\ell$.} and normalizing to the full theory, as
shown in Table~\ref{tab:fiterrors}.

We choose to parameterize the shape of the spectrum by the (normalized) first
and second derivatives at $x=1$, specifically $D'$ and $D''$ where
\begin{equation}
D(x) \simeq D_1 \left( 1 + D' (x-1) + {D''\over 2} (x-1)^2 \right)
\label{eq:quadspec}
\end{equation}
Note that $D'$ and $D''$ are $1/D_1$ times the first and second derivatives of
$D(x)$ at $x=1$.  The overall normalization is $D_1=110C_{10}$.
Below we quote the normalization as $D_1$ or $C_{10}$, for each $(D',D'')$
pair, and the goodness of fit by the relative likelihood of that shape
compared to a featureless, $n=1$, Sachs-Wolfe spectrum: $(D',D'')=(0,0)$.

\begin{figure}[t]
\centerline{\epsfxsize=3in\epsfbox{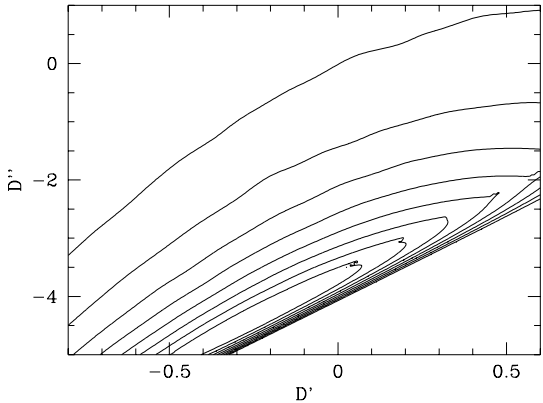}}
\caption{The likelihood $L(D',D'')$ is plotted for power
spectra given by Eq.~(18).  Likelihoods are normalized
so that a flat power spectrum ($D'=D''=0$) has $L=1$, and the contours
range from 1 to 8.}
\label{fig:quadlikely}
\end{figure}

In terms of this parameterization we find that the following fitting function
\begin{eqnarray}
\ln L &=&  -0.01669       + 1.19895D'  \\ \nonumber
      & &  - 0.83527D'^2  - 0.43541D'' \\ \nonumber
      & &  - 0.03421D'D'' + 0.01049D''^2
\label{eq:likelyfit}
\end{eqnarray}
describes the likelihood function with an error of 6\% in L over the
range $-0.6\le D'\le 0.6$ and $-1\le D''\le 3$.
The peak likelihood is at $(D',D'')=(-0.1,-4)$, which is outside the range
of the fitting function, and the likelihood at peak is 8.9 times that for a
flat spectrum.  The likelihood is plotted as a function of $D'$ and $D''$
in Fig.~\ref{fig:quadlikely}.

Similarly, the best-fitting value of $C_{10}$ is well approximated by
\begin{eqnarray}
10^{11}C_{10} &=&   0.64575      + 0.02282D' \\ \nonumber
              & & + 0.01391D'^2  - 0.01819D''\\ \nonumber
              & & - 0.00646D'D'' + 0.00103D''^2
\label{eq:c10fit}
\end{eqnarray}
with an error of 1\% over the same range.
The $1\sigma$ statistical uncertainty in $C_{10}$ is 13.8\%.
Uncertainties in rms quantities are therefore 7\%.\footnote{
These uncertainties are purely statistical.  There appears to be
an additional few-percent systematic uncertainty, based on the
fact that different pixelizations of the maps lead to slightly
different normalizations (\cite{Goretal96}).  One possible
estimate of the uncertainty might therefore be the quadrature sum
of 7\% and, say, 3\%, or 7.6\%.  However, since systematic errors
are generally highly non-Gaussian, a simple quadrature sum may
underestimate the uncertainty.  A conservative upper bound on
the uncertainty would be the sum of statistical and systematic
errors, or 10\%.  For a more detailed discussion of systematic
errors, see \cite{Goretal96,GorOpen,KogSys} and references therein.
}

\begin{figure}[t]
\centerline{\epsfxsize=3in\epsfbox{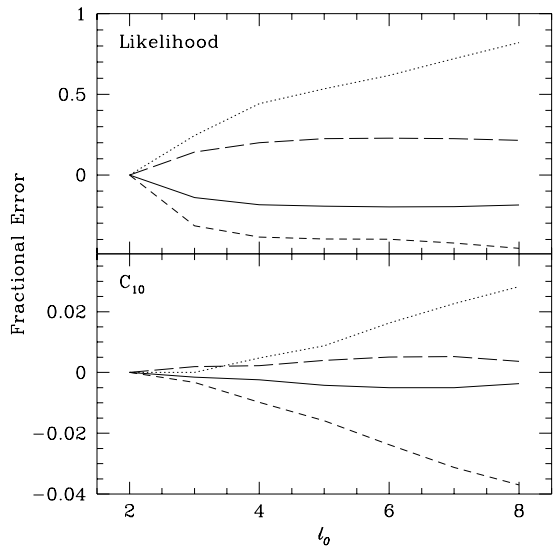}}
\caption{The error in the normalization and the likelihood that arises if a
theory which is quadratic in $\ell(\ell+1)C_\ell$ for $\ell\ge \ell_0$ but
constant for $\ell<\ell_0$ is approximated by the pure quadratic using the
fitting functions Eqs.~(19) and (20).
The underlying quadratics are taken to be $(D',D'')=$(-0.2,-1; solid),
(-0.2,1; dotted), (0.2, -1; short-dashed), (0.2,1; long-dashed).}
\label{fig:errors}
\end{figure}

Several popular theories (e.g.~open and $\Lambda$CDM models)
have radiation power spectra which are not well fit by quadratics all the
way down to the quadrupole (e.g.~Figs.~\ref{fig:SW}, \ref{fig:cl_omega}).
This is especially true if a tensor component is added since the $\ell=2$
mode is high in the tensor spectrum.
While the best-fitting normalization is not too sensitive to the assumption
of a quadratic over the entire range, this is not true of the likelihood.
We quantify in Fig.~\ref{fig:errors} and in Table~\ref{tab:fiterrors} what
error in the likelihood function and the normalization arise from
approximating several spectra by quadratics.
The quadratic approximation causes errors of consistently less than
1\% in the normalization $C_{10}$; these errors are
completely negligible in comparison with statistical and systematic
uncertainties.  The errors in the likelihood are somewhat larger, but
still not enough to significantly affect our conclusions.

\begin{table*}
\begin{tabular}{cccccc|ccccccc}
 GW & $\Omega_0$& $\Omega_\Lambda$ & $n$ & $\Delta C_{10}$(\%) &
  $\Delta{\cal L}$(\%) &
 GW & $\Omega_0$& $\Omega_\Lambda$ & $n$ & $\Delta C_{10}$(\%) &
  $\Delta{\cal L}$(\%) \\ \hline
  N & 0.20 & 0.80 & 1.00 &  0.5 &  -6 &   N & 0.20 & 0.80 & 1.10 &  0.8 & -18\\
  N & 0.30 & 0.00 & 0.90 & -1.2 & -14 &   N & 0.30 & 0.00 & 1.00 & -0.8 & -19\\
  N & 0.30 & 0.00 & 1.10 & -0.3 & -12 &   N & 0.30 & 0.70 & 1.00 &  0.7 & -10\\
  N & 0.30 & 0.70 & 1.10 &  0.7 & -17 &   N & 0.40 & 0.60 & 1.00 &  0.6 & -13\\
  N & 0.40 & 0.60 & 1.10 &  0.5 & -15 &   N & 0.50 & 0.50 & 0.90 &  0.7 &  -7\\
  N & 0.50 & 0.50 & 1.00 &  0.8 & -14 &   N & 0.50 & 0.50 & 1.10 &  0.6 & -12\\
  N & 0.60 & 0.00 & 0.90 &  0.8 &  -4 &   N & 0.60 & 0.00 & 1.00 &  0.6 &   4\\
  N & 0.60 & 0.40 & 0.90 &  0.7 &  -9 &   N & 0.60 & 0.40 & 1.00 &  0.7 & -12\\
  N & 0.60 & 0.40 & 1.10 &  0.5 & -15 &   N & 0.70 & 0.30 & 0.90 &  0.6 &  -7\\
  N & 0.70 & 0.30 & 1.00 &  0.6 &   0 &   N & 0.80 & 0.20 & 0.90 &  0.7 &  -9\\
  N & 0.80 & 0.20 & 1.00 &  0.6 &  -9 &   N & 0.90 & 0.10 & 0.90 &  0.6 &  -9\\
  N & 0.90 & 0.10 & 1.00 &  0.6 & -10 &   N & 0.90 & 0.10 & 1.10 &  0.3 & -14\\
  N & 1.00 & 0.00 & 0.90 &  0.6 &   4 &   N & 1.00 & 0.00 & 1.00 &  0.6 &  -6\\
  Y & 0.30 & 0.70 & 0.80 & -0.0 &  20 &   Y & 0.50 & 0.50 & 0.80 &  0.2 &  11\\
  Y & 0.60 & 0.40 & 0.80 &  0.1 &  21 &   Y & 0.70 & 0.30 & 0.80 &  0.1 &  31\\
  Y & 0.80 & 0.20 & 0.80 &  0.1 &  19 &   Y & 0.90 & 0.10 & 0.80 &  0.1 &  13\\
  Y & 1.00 & 0.00 & 0.80 &  0.0 &  34 &   Y & 0.50 & 0.50 & 0.90 &  0.5 &   5\\
  Y & 0.60 & 0.40 & 0.90 &  0.5 &   7 &   Y & 0.80 & 0.20 & 0.90 &  0.5 &  20\\
  Y & 0.90 & 0.10 & 0.90 &  0.4 &  13 &   Y & 1.00 & 0.00 & 0.90 &  0.3 &  11
\end{tabular}
\caption{The percentage error introduced by parameterizing theories as
quadratics.  We have compared the normalization and likelihood for 50
$\Lambda$CDM and OCDM theories with $0.8<n<1.1$ and $0.2\le\Omega_0\le1$ as
computed using Eqs.~(19,20) and directly computed from the data using the
numerical power spectrum.
Those theories where using the quadratic approximation induces an
error of $>0.5\%$ in $C_{10}$ or $>10\%$ in likelihood are shown above.
The column labeled GW indicates whether a component of gravity waves was
included, as discussed in the text.}
\label{tab:fiterrors}
\end{table*}

\begin{figure}[t]
\centerline{\epsfxsize=3in\epsfbox{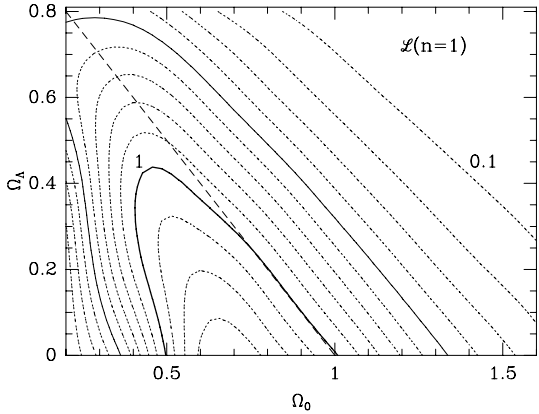}}
\caption{The likelihood function is shown for a family of CDM models with
$n=1$.  The likelihoods are normalized so that a flat power spectrum has $L=1$,
and the contours are separated by $0.1$ in $L$.  The straight dashed
line shows the spatially flat models.}
\label{fig:like_om_lam}
\end{figure}

\begin{figure}[t]
\centerline{\epsfxsize=3in\epsfbox{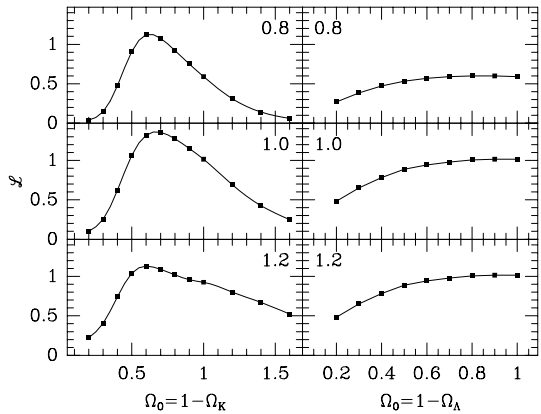}}
\caption{Likelihood as a function of $\Omega_0$ for CDM models with
zero cosmological constant (left) and zero spatial curvature (right).  The
spectral index $n$ increases from 0.8 to 1.2 from top to bottom.
The likelihoods are normalized so that a flat spectrum has $L=1$.}
\label{fig:like_om}
\end{figure}

The constraints on models from the shape of the {\sl COBE} power
spectrum are not strong.
Fig.~\ref{fig:like_om_lam} shows the likelihood function for a family of
CDM models in which the density and the cosmological constant are allowed
to vary while the spectral index is held fixed.
Fig.~\ref{fig:like_om} shows the likelihood function for models with
$\Omega_\Lambda=0$ and models with $\Omega_0+\Omega_\Lambda=1$ for three
values of the spectral index.  

The likelihoods in these figures
were computed using the quadratic approximation (\ref{eq:quadspec})
to the power spectrum rather than the exact power spectrum.  The 
fit to a quadratic is good to $\sim3\%$ for $\ell\ge 3$, but in
the worst cases the error in $C_2$ can be as much as 20\%.  This
causes errors of approximately 10-20\% in the likelihoods in the
worst cases but
does not change the general features of the plots.

It is clear from these figures that low values of $\Omega_0$ are disfavored,
especially in the open models.  To be more specific, the 95\% credible
Bayesian lower limit on $\Omega_0$ is 0.13 in a scale-invariant ($n=1$)
$\Lambda$CDM model, and the 95\% credible region for scale-invariant open
models is $0.3<\Omega_0<1.6$.
Both of these limits were computed using a uniform prior in $\Omega_0$.  In
the case of the cosmological constant models, the prior is of course nonzero
only when $0<\Omega_0<1$.

The low likelihood for low-$\Omega_0$ open models comes largely from the 
shape of the power spectrum at low $\ell$.  In fact, a significant portion
of our rejection power for these models comes from the quadrupole information.
If we exclude the quadrupole from consideration, we find, in agreement with
(\cite{GorOpen}), that we have no 95\%-confidence lower limit on $\Omega_0$
in a scale-invariant open model.
In the case of $\Lambda$ models, excluding the quadrupole weakens
the lower limit on $\Omega_0$ to about 0.08.

\section{Normalization of the Matter Spectrum} \label{sec:matter}

\subsection{Notation}

Until now we have talked about the normalization of the radiation power
spectrum on large angular scales.  Any theory of structure formation predicts
a definite ratio for the normalization of the radiation and matter power
spectra.  For this reason the {\sl COBE} DMR detection of CMB anisotropies
(\cite{Smoot}) allows us to directly normalize the potential fluctuations
at near-horizon scales.

For the matter density perturbations, the large-scale structure (LSS) 
data are usually expressed in
terms of the power spectrum $P(k)\equiv \left| \delta_k \right|^2$, where
$\delta_k$ is the Fourier transform of the fractional density perturbation
\begin{equation}
\delta_k\equiv\delta(|{\bf k}|)=
\int d^3x\ {\delta\rho\over\rho}\left({\bf x}\right) e^{i{\bf k}\cdot{\bf x}}.
\end{equation}
Since standard models postulate Gaussian fluctuations, specifying the power
spectrum completely determines the properties of the fluctuations.  As
the model has no preferred direction, the spectrum depends only on the
magnitude of ${\bf k}$.
Another measure of $P(k)$ that is often used is the contribution to the
mass variance per unit interval in $\ln k$, denoted $\Delta^2(k)$, which
has the virtue of being dimensionless:
\begin{equation}
\Delta^2(k)\equiv {d\sigma^2_{\rm mass}\over d\ln k} =
{k^3\over 2\pi^2}P(k).
\end{equation}
The normalization of $P(k)$ is frequently quoted in terms of
\begin{equation}
\sigma_8^2 \equiv \int_0^\infty {dk\over k}\ \Delta^2(k)\,
 \left( {3j_1(kr)\over kr}\right)^2,
\label{eqn:sigma8}
\end{equation}
with $r=8\,\hMpc$, which measures the variance of fluctuations in spheres
of radius $8\,\hMpc$.
Using the Press-Schechter or peak-patch methods, its value can be 
inferred from the abundance of clusters
(\cite{BonMye,WEF,Car94,VL,BonMye96})
to be $\sigma_8\simeq0.5$--0.8,
with some $\Omega_0$ dependence.  Specifically (\cite{VL}) find
\begin{equation}
\sigma_8\simeq(0.6\pm0.1)\Omega_0^{-\alpha}\ ,
\end{equation}
with $\alpha\simeq0.4$ for open CDM and $\alpha\simeq0.45$ for $\Lambda$CDM.
(More accurate fits plus a discussion of the uncertainty as a function of
$\Omega_0$ can be found in their paper.)
These values are consistent with those inferred from large-scale flows
(\cite{Dek94,StrWil}) and direct observations of galaxies (e.g.~\cite{Lov92}).
Note that for very low $\Omega_0$, this implies that optical galaxies become
anti-biased (i.e.~$b\equiv\sigma_8^{\rm gal}/\sigma_8 < 1$).
 
\begin{figure}[t]
\centerline{\epsfxsize=3in\epsfbox{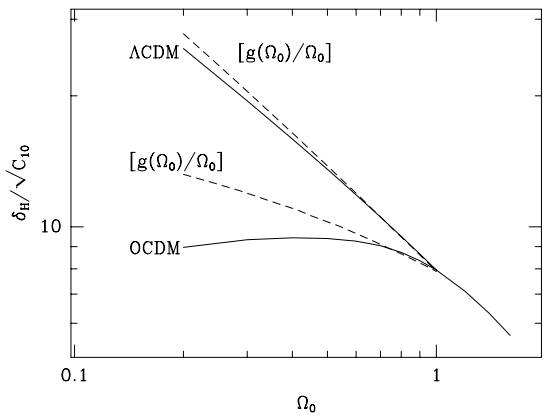}}
\caption{Relation between the normalization of the radiation power spectrum
(e.g.~$C_{10}$) and the matter power spectrum (e.g.~$\dH$) as a function
of $\Omega_0$ for CDM models with scale-invariant scalar perturbations.}
\label{fig:dhc10}
\end{figure}

While $\sigma_8$ has been the traditional means of quoting the normalization,
{\sl COBE} probes scales near the horizon size today, so we find it
cleanest to quote the normalization in terms of the amplitude of the mass
or potential fluctuations at {\it large\/} scales (small $k$).
Specifically we use $\dH$, the density perturbation at horizon-crossing,
which is defined through (see e.g.~\cite{LidLyt93})
\begin{equation}
\Delta^2(k) = {k^3 P(k)\over 2\pi^2}
            = \dH^2 \left( {k\over H_0}\right)^{3+n} T^2(k),
\label{eqn:delhdef}
\end{equation}
with $T(k)$ the transfer function describing the processing of the initial
fluctuations, which we take to be a power law with index $n$.  The relation
of the large-scale matter fluctuation amplitude, $\dH$, to the large-angle
radiation anisotropy amplitude, $\sqrt{C_{10}}$, is shown in
Fig.~\ref{fig:dhc10} and will be discussed more in
\S\S\ref{sec:flat}--\ref{sec:open}.
In the matter-dominated, critical density, Sachs-Wolfe approximation the
proportionality constant in Eq.~(\ref{eqn:swcl}) is simply $(\pi^2/8)\dH^2$.

\begin{figure}[t]
\centerline{\epsfxsize=3in\epsfbox{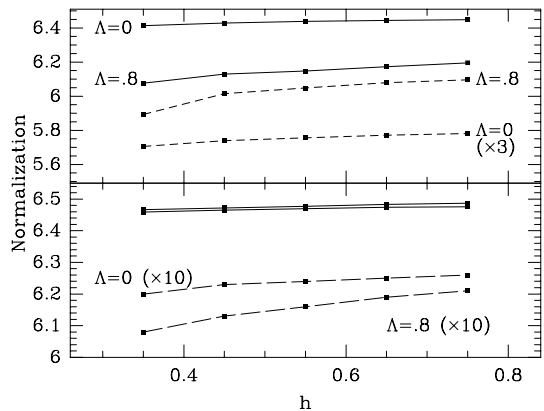}}
\caption{The power spectrum normalization for scalars (top panel) and
tensors (bottom panel) plotted as a function of the Hubble parameter for
two flat models.  Solid lines in both panels indicate $10^{12}C_{10}$ for
the best-fitting scale-invariant CDM model while dashed lines indicate $\dH$
(top panel) or $A_T$ (bottom panel).}
\label{fig:norm_vs_h}
\end{figure}

\begin{figure}[t]
\centerline{\epsfxsize=3in\epsfbox{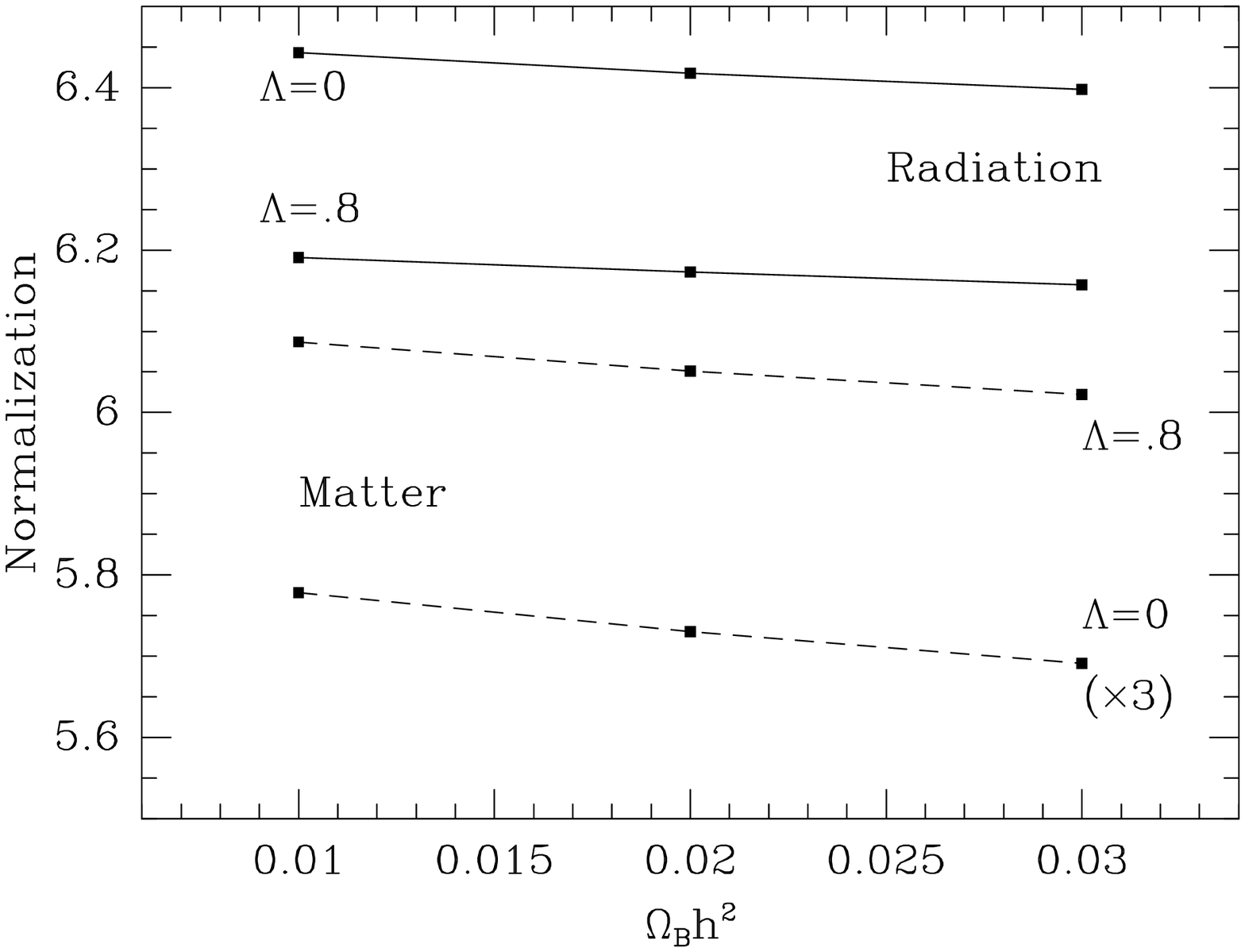}}
\caption{The power spectrum normalization for scalar perturbations versus
$\Omega_{\rm B}h^2$ for two flat models.  Solid lines indicate $10^{12}C_{10}$
for the best-fitting scale-invariant CDM model while dashed lines indicate
$\dH$.  The sensitivity of the tensors is similarly small.}
\label{fig:norm_vs_bbn}
\end{figure}

We find to very good approximation that $\dH$ as determined by {\sl COBE} is
independent of both $h$ and $\Omega_{\rm B}$
(see Figs.~\ref{fig:norm_vs_h}, \ref{fig:norm_vs_bbn}; for the open models
the sensitivity to $h$ is similar to the $\Omega_0=1$ case in
Fig.~\ref{fig:norm_vs_h} while the sensitivity to $\Omega_{\rm B}h^2$ is less
than in Fig.~\ref{fig:norm_vs_bbn}), although it will depend on $\Omega_0$ and
$\Omega_\Lambda$.  For definiteness we use $\Omega_{\rm B}h^2=0.0125$ and
$h=0.75$.  For high $\Omega_{\rm B}h^2$ and low $h$ one can scale $\dH$ using
Figs.~\ref{fig:norm_vs_h}, \ref{fig:norm_vs_bbn}.

The value of $\dH$ is related to the common inflationary amplitude of the
scalar perturbations $A_S$ (\cite{TurWhi,cobeinf}) by
$A_S\equiv\Omega_0/g(\Omega_0)\times\dH$, where $g(\Omega)$ is the growth
factor discussed in \S\ref{sec:flat}.
When we need to quote the normalization of the tensor (or gravity wave)
spectrum independently of the scalars, we shall normalize them to $A_T$,
where at large scales (\cite{TurWhi})
\begin{equation}
{d\Omega_{GW}\over d\ln k} \rightarrow {75\over 32}
        A_T^2 \left( {k\over H_0} \right)^{-2+n_T}
\end{equation}

Given $\delta_{\rm H}$, the value of $\sigma_8$ can be calculated using
Eq.~(\ref{eqn:sigma8}).
This will introduce an additional dependence on $n$, $\Omega_0$, $h$ and
the dark matter content of the universe, e.g.~$\Omega_{\rm HDM}$.  We
discuss this in \S\ref{sec:transfer}.

\subsection{Critical Density Models}

In the simplest picture, in which large-angle CMB anisotropies come 
purely from potential fluctuations on the last scattering surface, the 
relative normalization of the CMB and matter power spectrum today is 
straightforward (e.g.~\cite{Efs90,WSS}).
In a critical-density universe the potentials are constant in time.  If we
also assume adiabatic fluctuations, the temperature anisotropy is simply
one third of the potential fluctuation, which is related to the density
perturbation by the Poisson equation (see \cite{WhiHu} for a pedagogical
derivation).

For scale invariant scalar fluctuations in a CDM dominated universe with
$h=0.5$ and $\Omega_{\rm B}=0.05$ one obtains
$\delta_{\rm H}=1.92\times 10^{-5}$.
Over the range of scales probed by {\sl COBE} most models predict spectra
which are well approximated by power-laws, so we will allow ourselves to
depart from strictly scale-invariant models but keep a power-law initial
fluctuation spectrum.
The results for tilted ($n\ne1$) models are a special case of
Eq.~(\ref{eqn:dhlambda}).

In addition to scalar density fluctuations one must consider the possible
contribution of gravity waves (tensors) to the {\sl COBE} fluctuations.
If this contribution is non-negligible then the inferred matter fluctuations
are correspondingly lower.
Conventionally, this is defined in terms of the ratio of tensor to scalar
contribution to the quadrupole: $C_2^{(T)}/C_2^{(S)}$, also written as $T/S$.
If the inflationary model is specified then this quantity is calculable, and
is related to the tensor spectral index.
We shall treat two cases here, that where $T/S=0$ (corresponding to models
with a ``low'' scale of inflation, e.g.~\cite{Adaetal,LytSte96}) and that for
which
\begin{equation}
  {A_T^2\over A_S^2} = {1-n\over 3-n}
\end{equation}
which corresponds to power-law inflation (\cite{LucMat,LytSte92}).
In the $\Omega_0=1$ and $n\simeq1$ limit the above equation reduces to
the more familiar $T/S=7(1-n)$ (\cite{LidLyt,Davetal,Crietal,SteLyt}).
More details on the normalization of inflationary models with tensors,
and treatment of a wider class of models, can be found in (\cite{cobeinf}).
For a short discussion of the expectations for the amplitude of the tensor
spectrum from the point of view of particle physics models see
(\cite{Lyt}), and for a review of inflaton potential reconstruction and
further relations between observables see (\cite{Lidetal}).

\begin{figure}[t]
\centerline{\epsfxsize=3in\epsfbox{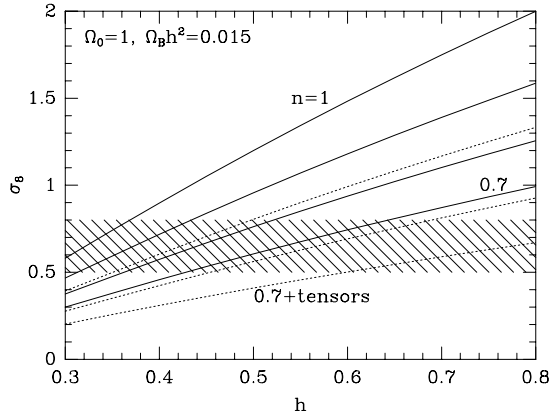}}
\caption{The {\sl COBE}-normalized small-scale fluctuation amplitude
$\sigma_8$ is plotted for a family of critical CDM models.
Solid lines are models with no tensor contribution, and dashed lines
have $T/S=7(1-n)$.  The hatched region shows the observed value of
$\sigma_8$ (see text).}
\label{fig:sig8crit}
\end{figure}
 
Fig.~\ref{fig:sig8crit} shows the {\sl COBE}-normalized value of $\sigma_8$
for a family of flat CDM models with $\Omega_0=1$.
While standard $n=1$ CDM predicts an unacceptably large value of $\sigma_8$
(unless the Hubble constant is extremely low), even a small amount of tilt
is sufficient to bring the prediction in line with the observations.

\subsection{Flat Models} \label{sec:flat}

There are several additional complications in the case when $\Omega_0\ne1$.
Let us first consider models with vanishing spatial curvature, but low
$\Omega_0$.  So we assume that there is a contribution from a cosmological
constant which restores spatial flatness: $\Omega_\Lambda=1-\Omega_0$.
In $\Lambda$CDM models the fluctuations stop growing when
$z=z_\Lambda\sim (\Omega_0^{-1}-1)^{1/3}$ for $\Omega_0\ll 1$,
so the overall growth from $z\sim10^3$ until the present is suppressed
by (e.g.~\cite{CPT})
\begin{equation}
g(\Omega) = {5\over 2} \Omega \left[ {1\over 70} + {209\Omega\over 140}
        - {\Omega^2\over 140} + \Omega^{4/7} \right]^{-1}  .
\end{equation}
In addition, the potential fluctuations are reduced by $\Omega_0$.
Thus in terms of the power spectrum, $P(k)$, we expect for fixed {\sl COBE}
normalization that $P(k)\propto(g(\Omega_0)/\Omega_0)^2\sim\Omega^{-1.54}$,
as has been pointed out by (\cite{Pee84,EBW}).
Hence for a fixed {\sl COBE} normalization the matter fluctuations today are
{\it larger\/} in a cosmological constant model than a critical density model.

However, the $g(\Omega_0)/\Omega_0$ behavior is not the only effect which
occurs in low-$\Omega_0$ universes.
Due to the fact that the potentials decay at $z_\Lambda$, there is another
contribution to the large-angle CMB anisotropy measured by {\sl COBE}.
On large angular scales, i.e., $\ell\la 10$ (\cite{KofSta,HuWhi}),
the blueshift of a photon falling into a potential well is not entirely
canceled by a redshift when it climbs out, the potential having decayed
during transit.
This leads to a net energy change, which accumulates along the photon path,
often called the Integrated Sachs-Wolfe (ISW) effect to distinguish it from
the more commonly considered redshifting which has become known as the
Sachs-Wolfe effect
(both effects were considered in the paper of \cite{SacWol}).
Fig.~\ref{fig:dhc10} shows the relation between the CMB power spectrum
normalization $C_{10}$ and the matter power spectrum normalization
$\delta_{\rm H}$.
Note that the $g(\Omega_0)/\Omega_0$ scaling of $\delta_{\rm H}$ is
fairly accurate for $\Lambda$CDM models where $\Omega_\Lambda$ domination
occurs late, but less accurate for open models where curvature domination
occurs early.

A fit to the four-year {\sl COBE} data for flat models gives the
horizon-crossing amplitude
\begin{equation}
10^5\,\dH(n,\Omega_0) = 1.94\; \Omega_0^{-0.785-0.05\ln\Omega_0}
        \exp \left[ a\widetilde{n} + b\widetilde{n}^2 \right] \,,
\label{eqn:dhlambda}
\end{equation}
where $\widetilde{n}=n-1$,  $a=-0.95$ and $b=-0.169$ with no gravitational
waves, and $a=1$ and $b=1.97$ with power-law inflation gravitational waves.
This fit works to better than 3\% for $0.2<\Omega_0\leq1$ and $0.7\le n\le1.3$,
and again the statistical uncertainty is 7\%.
For the power-law inflation case the fit is restricted to $n\le1$.
For $\Omega_0\ge0.3$ there is less than a 5\% correction to the
$g(\Omega_0)/\Omega_0$ scaling from the ISW component
(see Fig.~\ref{fig:dhc10}).

\begin{figure}[t]
\centerline{\epsfxsize=3in\epsfbox{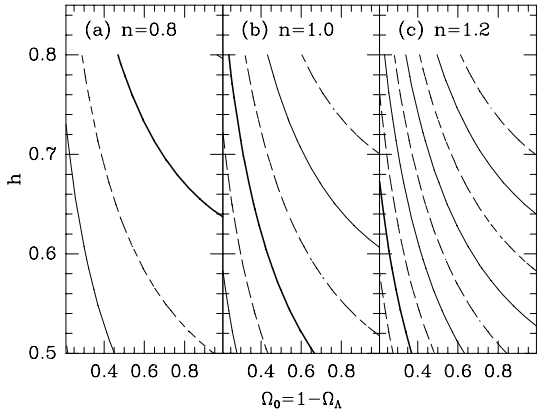}}
\caption{Contours of $\sigma_8$ are plotted for a family of flat CDM models,
with $n=0.8$, 1.0 and 1.2.  Contours are spaced by 0.25 with the thick solid
contour indicating $\sigma_8=1$.
In all cases $\Omega_{\rm B}h^2=0.015$ and we have used the small-scale
approximation to the transfer function.}
\label{fig:sig8flat}
\end{figure}
Going to smaller scales, Fig.~\ref{fig:sig8flat} shows
the
{\sl COBE}-normalized value of $\sigma_8$ for a variety of flat CDM models.

\subsection{Open Models} \label{sec:open}

If we again consider models with low $\Omega_0$ but now do not introduce a
cosmological constant to restore spatial flatness, we find that keeping track
of the ISW effect is extremely important.
Even though the calculations become more technically challenging in an open
model, it turns out that the largest effect is simple to understand.  Since
full matter domination occurs so far after last-scattering and curvature
domination occurs so early, the gravitational potentials are almost always
evolving ($\Phi$ is only constant when the universe is fully matter dominated).
For this reason, for the $\Omega_0$ of interest in structure formation, the
ISW effect dominates the Sachs-Wolfe 
effect over the entire range of angular scales.
Since the temperature anisotropy is enhanced by the ISW contribution, the
fluctuations in the matter required to fit {\sl COBE} are lowered in an open
model (see Fig.~\ref{fig:dhc10}).

There are additional complications due to the curvature selecting a
scale in the universe.  This means that there is some ambiguity in defining
the initial power spectrum.  We can obtain some guidance from inflationary
open-universe models which predict a nearly scale-invariant spectrum of
curvature (or gravitational potential) perturbations,
related to density perturbations through the Poisson equation.
The density fluctuation spectrum corresponding to a power-law curvature
perturbation is (\cite{LytSte} or see \cite{WhiBun} and references therein)
\begin{equation}
P(k) \propto {(q^2+4)^2\over q(q^2+1)} k^{n-1}
\end{equation}
with $q^2\equiv k^2/(-K)-1$ and $-K=H_0^2(1-\Omega_0)$.

In specific open-universe inflationary models based on bubble nucleation
there are corrections to this assumption.  First $P(k)$ is modified by
factors which can enhance or suppress power as $q\rightarrow0$, typically
a factor between $\coth(\pi q/2)$ and $\tanh(\pi q/2)$.
Second there is a spectrum of ``discrete'' modes with $k^2<1$ which can
add power at small $\ell$, and the fluctuations from the bubble wall can
give rise to perturbations in the open universe also (for a readable
introduction and overview of open inflationary models, see \cite{Coh}).
Furthermore, no calculations have yet been done to asses how good a power-law
the initial spectrum is supposed to be in the individual models.
Finally there is the possibility that a spectrum of gravitational waves is
produced, which should occur if the energy scale of inflation is close to the
GUT scale.  In short, the theoretical predictions for the power spectra
of such models are still quite uncertain.

To avoid detailed modeling of a particular inflationary scenario, we will
deal with simple power-law spectra of density perturbations here.
For the range of $\Omega_0$ of interest the $\coth(\pi q/2)$ and
$\tanh(\pi q/2)$ terms do not affect the normalization (\cite{YamBun}).
{}From the point of view of providing a fit to the large-scale structure and
CMB data, it appears that low-$\Omega_0$ models with a slight spectral tilt
$n\sim1.1$, no super-curvature or bubble wall modes and no tensor component
will fit the data best (\cite{WhiSil}), thus this assumption combines
simplicity with current prejudice.
A fit to the four-year {\sl COBE} data for such models gives the
horizon-crossing amplitude
\begin{eqnarray}
10^5\,\dH(n,\Omega_0) \kern-0.1in  & = & \kern-0.1in
	1.95\; \Omega_0^{-0.35-0.19\ln\Omega_0-0.17\widetilde{n}}
        \label{eqn:dhopen} \\ \nonumber
&&      \exp \left[ -\widetilde{n} -0.14\widetilde{n}^2 \right] \,.
\end{eqnarray}
This fit\footnote{After this paper was submitted (\cite{GorOpen}) also
considered open CDM models fit to the 4-year data.  Where our results overlap
there is excellent agreement: typically $\sim1\%$ in $\sigma_8$ for the range
of $\Omega_0$ of interest in structure formation.} works to better than 3\%
for $0.2<\Omega_0\le1$ and $0.7\le n\le1.2$, and again the statistical
uncertainty is 7\%.  This $\dH$ can be used in Eq.~(\ref{eqn:delhdef}) for
$k\gg\sqrt{-K}$, but note that the behavior of Eq.~(\ref{eqn:dhopen}) outside
the range $0.2<\Omega\le1$ is pathological.

\begin{figure}[t]
\centerline{\epsfxsize=3in\epsfbox{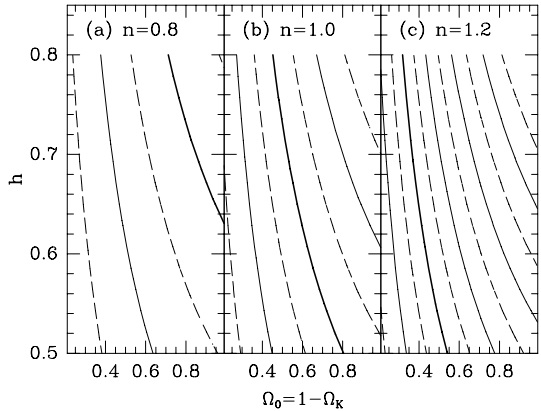}}
\caption{Contours of $\sigma_8$ are plotted for a family of open CDM models.
with $n=0.8$, 1.0 and 1.2.  Contours are spaced by 0.25, increasing to the
top right, with the thick solid contour indicating $\sigma_8=1$.
In all cases $\Omega_{\rm B}h^2=0.015$ and we have used the small-scale
approximation to the transfer function.}
\label{fig:sig8open}
\end{figure}

Note that the large difference in the scaling with $\Omega_0$ between these
models and flat models means that the {\sl COBE} normalization can distinguish
open and flat low-density models.
The addition of tensor or supercurvature modes would serve only to further
suppress the power on small scales in the open models, exacerbating the
difference.
This means that {\it open models can be distinguished from flat models\/}
using the {\sl COBE} normalization.
Fig.~\ref{fig:sig8open} shows the {\sl COBE}-normalized value of $\sigma_8$
for a variety of open CDM models.  For low $\Omega_0$ the scale-invariant
models whose shape fits the galaxy clustering data tend to underpredict the
cluster abundance (i.e.~$\sigma_8$) which is the basis of the claim above that
models with $n>1$ and no ``extra'' CMB fluctuations are preferred.

Typically cosmological models are assumed either to be spatially flat or to
have vanishing cosmological constant.  However, there is no reason in principle
not to consider models with both curvature and $\Lambda$.
The {\sl COBE} normalization for scale-invariant models with arbitrary
$\Omega_0$ and $\Omega_\Lambda$ is fit by
\begin{eqnarray}
10^5\delta_{\rm H}\kern-0.1in  & = & \kern-0.1in
  2.422 - 1.166e^{\Omega_0} + 0.800e^{\Omega_\Lambda} + 
  3.780 \Omega_0 - \nonumber \\
&& 2.267\Omega_0 e^{\Omega_\Lambda} + 
  0.487\Omega_0^2 + 0.561\Omega_\Lambda + \\
&& 3.392\Omega_\Lambda e^{\Omega_0} - 8.568\Omega_0 \Omega_\Lambda + 
   1.080\Omega_\Lambda^2 \nonumber
\end{eqnarray}
to an accuracy of 5\% for $0.2\le\Omega_0\le 1.6$ and 
$0\le\Omega_\Lambda\le 0.8$.
(For a discussion of such models, especially the closed ones, see
\cite{WhiScoClosed}.)

\subsection{Going to smaller scales} \label{sec:transfer}
 
Turning the {\sl COBE} normalization at the horizon scale into information
about clustering of the mass on smaller scales involves the transfer function
$T(k)$.  Several popular fitting functions exist for $T(k)$ in CDM and mixed
dark matter (MDM) models.
However, in order to exploit all of the accuracy of the {\sl COBE}
normalization, care must be taken with $T(k)$.
As pointed out by\footnote{The formula for $T(k)$ in \cite{PeaDod} contains
a typographical error of which readers should beware.}
(\cite{LidLyt93,PeaDod}) different parameterizations of the transfer function
can differ by relatively large amounts.
In addition, to obtain accuracy to better than the 10\% level, the effect of
baryon damping must be included (\cite{PeaDod,Sug,HuSugSml,Ma,WVLS}).
 
The MDM power spectrum can be obtained from the CDM power spectrum by
multiplying by an additional factor (\cite{PogSta,LLSSV,Ma}).  However, no
fully satisfactory parameterization of the CDM $T(k)$ has been published.
That is to say, no fitting function which is accurate to $\sim1\%$ over the
range $10^{-4}h\,{\rm Mpc}^{-1}\le k\le 1h\,{\rm Mpc}^{-1}$ for a wide range
of baryon densities (the scaling with the matter density and Hubble constant
is $\Omega_0h$ so can be included exactly).
 
\begin{figure}[t]
\centerline{\epsfxsize=3in\epsfbox{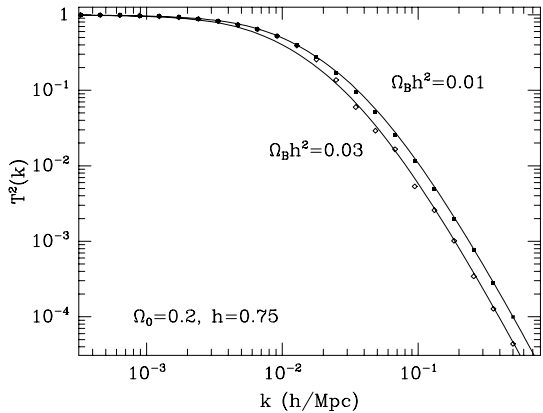}}
\caption{A comparison of the numerically evaluated transfer functions
(solid squares and open diamonds) with the small-scale approximation
discussed in the text (lines).  Note that even for extreme parameters the
approximation is quite good around $k\sim0.2 h\,{\rm Mpc}^{-1}$ where
$\sigma_8$ receives its main contribution.}
\label{fig:transfer}
\end{figure}
 
For the purposes of this paper, we shall compute only $\sigma_8$.  Thus
we need an approximation to $T(k)$ which is accurate on small scales.
We use the fitting function of (\cite{BBKS}),
\begin{eqnarray}
T_{{\rm CDM}}(q) & = & \frac{\ln \left(1+2.34q \right)}{2.34q} \times
        \\ \nonumber
& & \left[1+3.89q+(16.1q)^2+ \right. \\ \nonumber
& & \left. (5.46q)^3+(6.71q)^4\right]^{-1/4} \,,
\end{eqnarray}
with $q = k/h\Gamma$, and $\Gamma$ given by (\cite{HuSugSml}, Eq.~D-28, E-12).
As we show in Fig.~\ref{fig:transfer} this works well even in extreme cases, if
one averages over the small ripples inherent in the large $\Omega_B/\Omega_0$
transfer function.
The deviation for $k\ge 0.1h\,{\rm Mpc}^{-1}$ is less than a few percent,
though at larger scales it can be significant.  The approximation
$\Gamma=\Omega_0h\exp(-\Omega_B-\Omega_B/\Omega_0)$ of (\cite{Sug}) works
relatively well at large scales for low $\Omega_B/\Omega_0$ but can lead to
unacceptable errors in $T(k)$ outside of this range.

\section{Implications for Large-Scale Structure}
\label{sec:lss}

\begin{table*}[t]
\begin{center}
\begin{tabular}{llllllll}
\hline
 & $\Omega_0$ & $\Omega_\Lambda$ & $\Omega_\nu$ & $n$ & $h$
& $\Omega_{\rm B}h^2$ & $\sigma_8$ \\
\hline
standard CDM & 1.0 & 0.0 & 0.0 & 1.0 & 0.50 &  0.0125 &   1.22 \\
tilted CDM & 1.0 & 0.0 & 0.0 & 0.8 & 0.50 &  0.0250 &   0.72 \\
MDM & 1.0 & 0.0 & 0.2 & 1.0 & 0.50 &  0.0150 &   0.79 \\
$\Lambda$CDM & 0.4 & 0.6 & 0.0 & 1.0 & 0.65 &  0.0150 &   1.07 \\
Open CDM & 0.4 & 0.0 & 0.0 & 1.0 & 0.65 &  0.0150 &   0.64 \\
Low $h$ CDM &  1.0 & 0.0 & 0.0 & 1.0 & 0.35 & 0.0150 &   0.74\\
\hline
\end{tabular}
\end{center}
\caption{Small-scale normalizations for various models.}
\label{table}
\end{table*}

In principle, {\sl COBE} lets us constrain
models of large-scale structure in two independent ways: we
can use both the shape
of the power spectrum and the normalization.  In practice, though,
the shape constraints turn out to be quite weak for most popular
models, as we saw in \S\ref{sec:radiation} and Fig.~\ref{fig:like_om}.
We therefore focus on the implications of the power spectrum normalization.

Table~\ref{table} gives values of the small-scale density fluctuation
amplitude $\sigma_8$ for a selection of models.
The first line makes it clear that {\sl COBE}-normalized ``standard''
CDM ($n=1$, $h=0.5$, $\Omega_{\rm B}h^2=0.0125$) predicts significantly
too much small-scale power and is therefore ruled out.
However, any of several slight changes to the model can easily resolve this
inconsistency.  Perhaps the simplest solution is a slight tilt to the power
spectrum.  Inflationary models typically predict spectral indices slightly
less than one, and a value of $n$ of 0.8 or even less is quite natural in
such models.

We have discussed the difference in normalization between the open and
$\Lambda$CDM models in the previous section.
By reducing the density we reduce the small-scale fluctuation amplitude
as measured, say, by $\sigma_8$.  Since $\sigma_8$ is typically too large
in ``standard'' critical-density CDM models, such low-density models tend
to fare better.  As we saw above, the suppression of $\sigma_8$ is greater
in open models than in $\Lambda$CDM models.  These results are quantified
for some specific models in Table~\ref{table}.

Another obvious impact of the {\sl COBE} normalization is that it
differentiates between the classes of theories based on adiabatic
fluctuations (e.g.~inflation) and those based on isocurvature
fluctuations (e.g.~topological defect models).  In the adiabatic
models, the fluctuations in all the constituents have the same sign, so
an overdensity of photons is also an overdensity of baryons and cold dark
matter.  This means that photons from overdense regions (which are
hotter than average, $n\sim T^3$) must climb out of larger than
average potential wells, in the process losing energy.  In
isocurvature models it is ``cold'' photons which must climb out of
such potential wells.  Thus, in the adiabatic case the ``intrinsic''
temperature fluctuation and the gravitational redshifting effect
partially cancel, while in the isocurvature case they add.  This leads
to smaller temperature fluctuations for a given matter fluctuation in
the adiabatic case, or conversely larger matter
fluctuations when normalized to the temperature fluctuations observed
by {\sl COBE}.

The relative normalization of the matter and temperature fluctuations in any
particular model depends upon the details of that model, but the above
argument is general enough that we may expect it to hold in a large class
of models.  Since the adiabatic models predict roughly the right amount of
power when compared to large-scale structure measurements on smaller scales,
the generic isocurvature models predict too little.  These issues are dealt
with in more detail in \cite{WhiSco} and references therein.

Discussion of particular models of structure formation is beyond the scope
of this paper.  Analyses based on the four-year {\sl COBE} data
have been performed for critical-density models (\cite{WVLS}),
$\Lambda$CDM models (\cite{LLVW}), and open CDM models (\cite{GorOpen,WhiSil}).
The main difference between the four-year results and those based on analyses
of the two-year data (e.g., \cite{Goretal,StoGorBan,WhiBun}.  See also
\cite{WhiSco} and references therein.) is a reduction of $\sim10\%$ across the
board in the {\sl COBE} normalizations.  Part of this reduction is due to
Galactic contamination in the two-year analyses: the two-year data were
typically analyzed with a straight $20^\circ$ Galactic cut, as opposed to the
slightly more extensive ``custom cut'' (\cite{Ben96}) that has been applied
to the four-year data.

\section{Conclusions} \label{sec:conclusions}

The four-year {\sl COBE} DMR sky maps provide some of the most stringent
constraints on cosmological models.  Because the data are so powerful,
and because they are the best large-angle CMB data we are likely to have
for the next few years, it is important to extract as much information as
possible from them.  In particular, when using {\sl COBE} to normalize models
for large-scale structure formation, one should perform a full fit to the
data, rather than using a single number such as the root-mean-square
temperature fluctuation smoothed on some angular scale.

We have provided normalizations and likelihoods based on the four-year
DMR data for a broad class of cosmological models.  In conjunction with
small-scale measurements of the matter power spectrum such as $\sigma_8$,
these {\sl COBE} normalizations allow us to place significant constraints
on the parameters of various cosmological models.  Furthermore,
CMB data on smaller angular scales is improving rapidly in both
quality and quantity.  By comparing these data
with the estimates of the {\sl COBE} power spectrum, it will
soon be possible to severely constrain models on the basis
of the CMB angular power spectrum alone.

\acknowledgments

We wish to thank W.~Hu, A.~Liddle, D.~Scott and J.~Silk for useful
conversations and E.~Wright for supplying the software used
to make Fig.~\ref{fig:klmaps}.  EFB was supported by NASA.

\clearpage

%\twocolumn

\clearpage

%\onecolumn


\begin{thebibliography}{}
\bibitem[Abbott \& Wise~1984]{AbbWis}
Abbott, L.F. \& Wise, M.B.~1994, \apj, 282, L47.
\bibitem[Adams et al.~1993]{Adaetal}
 Adams, F.C., Bond, J.R., Freese, K., Frieman, J.A., Olinto, A.V., 1993
 Phys. Rev. D47, 426.
\bibitem[Banday et al.~1994]{Ban94}
 Banday, A. et al.~1994, \apj, 436, L99.
\bibitem[Banday et al.~1996]{Ban96}
 Banday, A. et al.~1996, preprint (astro-ph/9601065).
\bibitem[Bardeen et al.~1986]{BBKS}
 Bardeen J. M., Bond J. R., Kaiser N., \& Szalay A. S.~1986, \apj, 304, 15.
\bibitem[Bennnett et al.~1996]{Ben96}
 Bennett, C.L. et al.~1996, \apj, 464, L1.
\bibitem[Bond 1995]{Bond}
 Bond, J.R.~1995, Phys. Rev. Lett., 74, 4369.
\bibitem[Bond \& Efstathiou~1987]{BonEfs}
 Bond, J.R. \& Efstathiou, G.~1987, \mnras, 226, 655.
\bibitem[Bond \& Myers~1991]{BonMye}
 Bond, J. R. \& Myers, S., 1991, in ``Trends in Astroparticle Physics'',
 eg.~D.~Cline and R.~Peccei, Singapore, World Scientific, p.262.
\bibitem[Bond \& Myers~1996]{BonMye96}
 Bond, J. R. \& Myers, S., 1996, \apjs, 103, 63.
\bibitem[Bunn~1995]{Bunn}
 Bunn, E.F.~1995, Ph.D. thesis, University of California, Berkeley.
\bibitem[Bunn, Hoffman, \& Silk~1996]{BunHofSilk}
 Bunn, E.F., Hoffman, Y., \& Silk, J.~1996, \apj, 464, 1.
\bibitem[Bunn, Liddle \& White~1996]{cobeinf}
 Bunn, E. F., Liddle, A.R., \& White, M.~1996, Phys. Rev. D., in press
 (astro-ph/9607038).
\bibitem[Bunn, Scott \& White~1995]{BunScoWhi}
 Bunn, E.F., Scott D., \& White M.~1995, \apj, 441, L9.
\bibitem[Bunn \& Sugiyama~1995]{BunSug}
 Bunn, E. F. \& Sugiyama, N.~1995, \apj, 446, 49.
\bibitem[Bunn et al.~1994]{BunWhiSreSco}
 Bunn, E.F., White, M., Srednicki, M., \& Scott, D.~1994, \apj, 429, 1.
\bibitem[Carlberg et al.~1994]{Car94}
 Carlberg, R.~et al., 1994, J. R. Astron. Soc. Canada, 88, 39.
\bibitem[Carroll et al.~1992]{CPT}
 Carroll, S.M.,  Press, W.H., \& Turner, E. L.~1992, \araa, 30, 499.
\bibitem[Cohn~1996]{Coh}
 Cohn, J.D.~1996, in ``Microwave Background Anisotropies'', Proceedings of
 the XXXI Moriond Meeting, ed.~F.~Bouchet., (astro-ph/9606052).
\bibitem[Crittenden et al.~1993]{Crietal}
 Crittenden R., Bond J.R., Davies R.L., Efstathiou G., Steinhardt P.J.,
 1993, Phys. Rev. Lett. 71, 324.
\bibitem[Davis et al.~1992]{Davetal}
 Davis, R.L., Hodges H.M., Smoot G.F., Steinhardt P.J., \& Turner M.S.~1992,
 Phys Rev Lett, 69, 1856; erratum ibid 70, 1733.
\bibitem[Dekel~1994]{Dek94}
 Dekel, A.~1994, \araa, 32, 371.
\bibitem[Efstathiou~1990]{Efs90}
 Efstathiou, G., 1990, in ``Physics of the Early Universe'',
 ed.  J.A. Peacock, A.E. Heavens \& A.T. Davies, Adam Hilger,
 New York, p.$\,361$.
\bibitem[Efstathiou, Bond \& White~1992]{EBW}
 Efstathiou G., Bond J. R., \& White S. D. M.~1992, \mnras, 258, 1P.
\bibitem[Fixsen et al.~1996]{Fixsen}
 Fixsen, D.J., Cheng, E.S., Gales, J.M., Mather, J.C., Shafer, R.A.,
 \& Wright, E.L.~1996, preprint (astro-ph/9605054).
\bibitem[G{\'o}rski et al.~1995]{Goretal}
 G{\'o}rski, K. M., Ratra, B., Sugiyama, N., \&
 Banday, A. J.,~1995 \apj, 446, L67.
\bibitem[G\'orski et al.~1996]{Goretal96}
 G\'orski, K.M. et al.~1996a, \apj, 464, L11.
\bibitem[G\'orski et al.~1996b]{GorOpen}
 G\'orski, K.M. et al.~1996b, preprint (astro-ph/9608054).
\bibitem[Hinshaw et al.~1996]{Hin96}
 Hinshaw, G. et al.~1996, \apj, 464, L17.
\bibitem[Hu, Bunn \& Sugiyama~1995]{HuBS}
 Hu, W., Bunn, E. F., \& Sugiyama, N.~1995, \apj, 447, L59.
\bibitem[Hu, Scott \& Silk~1994]{HuScoSil}
 Hu, W., Scott, D., \& Silk, J.~1994, Phys. Rev. D, 49, 648.
\bibitem[HSSW]{HSSW}
 Hu, W., Scott, D., Sugiyama, N., \& White, M.~1995, Phys.Rev. D52, 5498.
\bibitem[Hu \& Sugiyama~1996]{HuSugSml}
 Hu, W. \& Sugiyama, N.~1996, \apj, in press, (astro-ph/9510117).
\bibitem[Hu \& White~1996]{HuWhi}
 Hu, W. \& White, M.~1996, A\&A, 315, 33, (astro-ph/9507060).
\bibitem[Kofman \& Starobinsky~1985]{KofSta}
 Kofman, L. \& Starobinsky, A. A.~1985, Sov. Astron. Lett., 9, 643.
\bibitem[Kogut et al.~1996a]{Kog96}
 Kogut, A. et al.~1996a, \apj, 464, L29.
\bibitem[Kogut et al.~1996b]{KogSys}
 Kogut, A. et al.~1996b, \apj, 470, 653.
\bibitem[Liddle \& Lyth~1992]{LidLyt}
 Liddle, A.R. \& Lyth, D.~1992, Phys. Lett. B291, 391.
\bibitem[Liddle \& Lyth~1993]{LidLyt93}
 Liddle, A.R. \& Lyth, D.~1993, Phys. Rep., 231, 1.
\bibitem[Liddle et al.~1996a]{LLSSV}
 Liddle, A.R. et al.~1996a, \mnras, 281, 531.
\bibitem[Liddle et al.~1996b]{LLVW}
 Liddle, A.R., Lyth, D.H., Viana, P.T.P., \& White, M. 1996b, \mnras,
 282, 281.
\bibitem[Lidsey et al. 1996]{Lidetal}
 Lidsey, J. et al.~1996, Rev. Mod. Phys., in press (astro-ph/9508078).
\bibitem[Lineweaver et al. 1994]{Line}
 Lineweaver, C.H. et al.~1994, \apj, 436, 452.
\bibitem[Loveday et al.~1992]{Lov92}
 Loveday, J., Efstathiou, G., Peterson, B.A., \& Maddox, S.J.~1992,
 \apj, 400, L43.
\bibitem[Lucchin \& Matarrese~1985]{LucMat}
 Lucchin F. \& Matarrese S.~1985, Phys Rev, D32, 1316.
\bibitem[Lyth~1996]{Lyt}
 Lyth D.H.~1996, preprint, hep-ph/9606387.
\bibitem[Lyth \& Stewart~1990]{LytSte}
 Lyth D.H., Stewart E.~1990, Phys. Lett. B252, 336.
\bibitem[Lyth \& Stewart~1992]{LytSte92}
 Lyth D.H., Stewart E.~1992, Phys. Lett. B274, 168.
\bibitem[Lyth \& Stewart~1996]{LytSte96}
 Lyth D.H., Stewart E.~1996, preprint, hep-ph/9606412
\bibitem[Ma~1996]{Ma}
 Ma, C.-P.~1996, \apj, 471, ? (astro-ph/9605198).
\bibitem[Peacock \& Dodds~1994]{PeaDod}
 Peacock, J.A. \& Dodds, S.J.~1994, \mnras, 267, 1020.
\bibitem[Peebles~1984]{Pee84}
 Peebles, P. J. E.~1984, \apj, 284, 439.
\bibitem[Peebles~1993]{Pee93}
 Peebles, P. J. E.~1993, ``Principles of Physical Cosmology'',
 (Princeton University Press, Princeton, New Jersey) \S21.
\bibitem[Pogosyan \& Starobinsky~1995]{PogSta}
 Pogosyan, D. \& Starobinsky, A.~1995, \apj, 447, 465.
\bibitem[Sachs \& Wolfe~1967]{SacWol}
 Sachs, R. K. \& Wolfe, A. M.~1967, \apj, 147, 73.
\bibitem[Scott, Silk \& White~1995]{SSW}
 Scott, D., Silk, J., \& White, M.~1995, Science, 268, 829.
\bibitem[Smoot et al.~1992]{Smoot}
 Smoot, G. et al.~1992, \apj, 396, L1.
\bibitem[Stompor, Gorski \& Banday 1995]{StoGorBan}
 Stompor, R., Gorski, K., \& Banday, A.~1995, \mnras, 277, 1225.
\bibitem[Strauss \& Willick~1995]{StrWil}
 Strauss, M.A. \& Willick, J.~1995, Phys. Rep., 261, 271.
\bibitem[Stewart \& Lyth~1993]{SteLyt}
 Stewart, E. \& Lyth, D.H., 1993, Phys. Lett. B302, 171.
\bibitem[Sugiyama 1995]{Sug}
 Sugiyama, N.~1995, Ap.J.Supp, 100, 281.
\bibitem[Tegmark \& Bunn~1995]{TegBun}
 Tegmark, M. \& Bunn, E.F. 1995~\apj, 455, 1.
\bibitem[Turner \& White~1996]{TurWhi}
 Turner, M. S. \& White, M.~1996, Phys. Rev. D53, 6822.
\bibitem[Viana \& Liddle~1996]{VL}
 Viana, P.T.P. \& Liddle, A.~1996, \mnras, 281, 323.
\bibitem[Vogeley \& Szalay~1996]{VogSza}
 Vogeley, M. S. \& Szalay, A. S.~1996, \apj, 465, 34.
\bibitem[White \& Bunn~1995]{WhiBun}
 White, M. \& Bunn, E.F.~1995, \apj, 450, 477.
\bibitem[White, Efstathiou \& Frenk~1993]{WEF}
 White, S.D.M., Efstathiou, G. \& Frenk, C.~1993, \mnras, 262, 1023.
\bibitem[White \& Hu~1996]{WhiHu}
 White, M. \& Hu, W.~1996, IAS preprint, (astro-ph/9609105).
\bibitem[White \& Scott~1996a]{WhiSco}
 White, M. \& Scott, D.~1996a, Comments on Astrophys., 18, 289,
 (astro-ph/9601170).
\bibitem[White \& Scott~1996b]{WhiScoClosed}
 White, M. \& Scott, D.~1996b, \apj, 459, 415.
\bibitem[White \& Silk~1996]{WhiSil}
 White, M. \& Silk, J.~1996, Phys. Rev. Lett., in press, (astro-ph/9608177).
\bibitem[White et al.~1994]{WSS}
 White, M., Scott, D., \& Silk, J.~1994, \araa, 32, 319.
\bibitem[White et al.~1996]{WVLS}
 White, M., Viana, P.T.P., Liddle, A. \& Scott, D.~1996, \mnras,
 in press (astro-ph/9605057).
\bibitem[Wright et al.~1994]{Wri94}
 Wright, E.L. et al.~1994, \apj, 420, 1.
\bibitem[Yamamoto \& Bunn~1996]{YamBun}
 Yamamoto, K. \& Bunn, E.F.~1996, \apj, 464, 8
\end{thebibliography}
\end{document}